\documentclass[twocolumn,showpacs,preprintnumbers, prd,
			 superscriptaddress,nofootinbib]{revtex4}

\usepackage{graphicx, fancybox, color}
\usepackage{amsmath,amssymb,bm}
\usepackage{braket}
\begin{document}

\title{Dynamics of in-medium quarkonia in SU(3) and SU(2) gauge theories}
\author{Yukinao Akamatsu}%
\email{akamatsu@kern.phys.sci.osaka-u.ac.jp}
\affiliation{%
 Department of Physics, Osaka University, Toyonaka, Osaka 560-0043, Japan
}%
\author{Masayuki Asakawa}%
\email{yuki@phys.sci.osaka-u.ac.jp}
\affiliation{%
 Department of Physics, Osaka University, Toyonaka, Osaka 560-0043, Japan
}%
\author{Shiori Kajimoto}
\email{kajimoto@kern.phys.sci.osaka-u.ac.jp}
\affiliation{%
 Department of Physics, Osaka University, Toyonaka, Osaka 560-0043, Japan
}%

\begin{abstract}
Decoherence dynamics of quarkonia is studied in the high-temperature deconfined phase of SU($N_c$) gauge theories.
In particular, we analyze the symmetry properties of SU($N_c$) stochastic potential model and find a novel ``event-by-event" symmetry for $N_c=2$ case, similar to the $G$-parity of hadronic systems.
This novel symmetry constrains the relation between diagonal and off-diagonal components of quarkonium density matrix, leaving the latter to be finite at late times.
We also present one-dimensional numerical simulation of the model, which indicates the usefulness of the complex potential simulations for the quarkonium survival probabilities in relativistic heavy-ion collisions, provided that the effect of dissipation can be neglected.
\end{abstract}

\date{\today}
\maketitle

\section{Introduction}\label{sec:intro}
Decoherence is a fundamental concept in quantum physics \cite{Schlosshauer:2019ewh}.
It explains the reason why we do not encounter macroscopic superposition states, such as {\it the Schr\"odinger's cat}, in our daily lives.
Essential observation is that any quantum mechanical system is almost inevitably coupled to environmental degrees of freedom.
Such coupling induces entanglement between the system and the environment and the interference in the macroscopic superposition disappears in an instant.
So, where there's an environment, there's decoherence.
In the era of quantum computing, it is crucial to avoid or minimize the effects of decoherence in quantum devices in order to prepare and maintain superposition states of the qubits.

More generally, decoherence is a gradual physical process in which a superposition state turns into a mixed state with classical probability.
Developments in the experimental techniques enable one to create mesoscopic superposition states and to measure their gradual decoherence.
Examples include photon coherent states in superposition in a photon cavity \cite{Brune:1996zz}, separated wave packets in ion traps \cite{myatt2000decoherence}, matter wave of a massive ${\rm C}_{70}$ molecule in Talbot-Lau interferometer \cite{Hornberger_2003}, superposition of superconducting flux qubits \cite{Chiorescu_2003}, and so on.

Recently, the decoherence of a quantum mechanical bound state of heavy quark pair, or quarkonium, is attracting attention \cite{Akamatsu:2020ypb}.
The environment for quarkonium is an extremely hot state of matter, called quark-gluon plasma (QGP) \cite{Yagi:985487}.
The QGP is a strongly coupled plasma composed of quarks and gluons, which are liberated from the hadrons, and has been vigorously investigated in relativistic heavy-ion collision experiments at Relativistic Heavy-Ion Collider (RHIC) and the Large Hadron Collider (LHC).
So far it is the most perfect fluid known in the nature with the shear viscosity to entropy density ratio $\eta/s \sim (1$-$2) /4\pi$ \cite{Bernhard:2019bmu} close to the conjectured universal lower bound of $1/4\pi$ by the gauge/gravity correspondence \cite{Kovtun:2004de}.

In the classic paper \cite{Matsui:1986dk}, Matsui and Satz showed that the quarkonium cannot survive the QGP environment when the color screening is so strong that the in-medium force cannot bind the heavy quark pair.
It was not until 2011 that the decoherence is recognized as another source of quarkonium dissociation in the QGP environment \cite{Akamatsu:2011se}.
In short, the color screening is a static medium effect while the decoherence is a dynamical one.

Theoretically, the dynamics of quarkonium in the QGP is studied in the framework of open quantum systems \cite{Akamatsu:2012vt, Rothkopf:2013kya, Akamatsu:2014qsa, Kajimoto:2017rel, Akamatsu:2018xim, Miura:2019ssi, Blaizot:2015hya, Blaizot:2017ypk, Blaizot:2018oev, Brambilla:2016wgg, Brambilla:2017zei, Brambilla:2019tpt, Escobedo:2020tuc, Brambilla:2020qwo, Brambilla:2021wkt, Omar:2021kra, Young:2010jq, Borghini:2011yq, Borghini:2011ms, Katz:2015qja, DeBoni:2017ocl, Yao:2018nmy, Sharma:2019xum}
(see also \cite{Rothkopf:2019ipj, Akamatsu:2020ypb, Yao:2021lus, Sharma:2021vvu} for recent reviews).
In particular, we adopt the regime of the quantum Brownian motion, in which the time scale of quarkonium $\tau_S$ is assumed to be much longer than the QGP correlation time $\tau_E$ \cite{BRE02, Caldeira:1982iu}.
For a rough estimate, we take $1/\tau_S$ to be the binding energy in the Coulomb potential with $V(r) = -C_F\alpha_s / r$ and $1/\tau_E$ to be the minimum (nonzero) Matsubara frequency:
\begin{align}
\tau_S \sim 4/M(C_F\alpha_s)^2,\quad
\tau_E\sim 1/2\pi T.
\end{align}
Here $M$ is the heavy quark mass ($M \approx 1.18 {\rm GeV}$ for charm and $M\approx 4.8 {\rm GeV}$ for bottom) and $C_F\alpha_s \approx 0.3\text{-}0.4$ is the coefficient of color Coulomb force in the phenomenological Cornell potential \cite{Bali:2000gf, Alford:2013jva}, and then
\begin{subequations}
\begin{align}
\tau_S({\rm bottomonia})&\sim [(0.11\text{-}0.19) {\rm GeV}]^{-1} ,\\
\tau_S({\rm charmonia}) &\sim [(0.027\text{-}0.047) {\rm GeV}]^{-1}.
\end{align}
\end{subequations}
Typically, $T\sim 0.2\text{-}0.4 {\rm GeV}$ for the QGP created in heavy-ion collisions, and thus the time scale hierarchy $\tau_S\gg \tau_E$ seems to hold\footnote{
Let us comment, however, that the time scale hierarchy $\tau_S\gg \tau_E$ for bottomonia is marginal without the factor $2\pi$ in $\tau_E$.
There is a different approach to derive and solve a classical kinetic theory describing the singlet quarkonia bound states as molecules by interpreting $\tau_S\lesssim \tau_E$ \cite{Yao:2017fuc, Yao:2018nmy, Yao:2018sgn, Yao:2020xzw}.
}.
In this regime, the decoherence of quarkonium proceeds by its coupling to color electric sector of the QGP and thus is determined by the properties of color electric fluctuations, which are encoded in the in-medium complex potential \cite{Akamatsu:2014qsa}.
In the dipole limit, the properties of color electric fluctuations are quantified by local coefficients such as the heavy quark momentum diffusion constant and the thermal dipole self-energy constant \cite{Brambilla:2016wgg, Brambilla:2017zei, Brambilla:2019tpt}.
The observed suppression pattern of bottomonia family $\Upsilon (n{\rm S})$ \cite{Adamczyk:2013poh, Adare:2014hje, Chatrchyan:2011pe, Chatrchyan:2012lxa, Khachatryan:2016xxp, Abelev:2014nua} in the heavy-ion collisions must contain such information of the QGP\footnote{
Interpretation of the charmonum data \cite{Adare:2006ns, Adare:2008sh, Adare:2011yf, Adamczyk:2013tvk, Abelev:2013ila, Adam:2016rdg} 
requires some care because the probability of charmonium regeneration from initially uncorrelated charm quark pair might not be negligible \cite{BraunMunzinger:2000px, Zhao:2011cv, Song:2011xi, Zhou:2014kka}, in particular at the LHC \cite{Abelev:2013ila, Adam:2016rdg}.
}.

In this paper, we show our theoretical analysis for the quarkonium decoherence in the deconfined phase of SU($N_c$) gauge theories at high temperatures.
Our formulation is based on SU($N_c$) stochastic potential model formulated in \cite{Akamatsu:2014qsa}.
Our previous paper \cite{Kajimoto:2017rel} was also motivated by quarkonium dissociation in the QGP,  but the numerical simulation was carried out for U(1) gauge theory (QED).
The extension to SU($N_c$) introduces the color internal space for quarkonia, the singlet and octet (or $N_c^2-1$ representation), and the color dependent potential force, which is attractive/repulsive for quarkonia in the singlet/octet.
The simulation results for the bound state occupation in a static and an expanding QGP supports the complex potential model \cite{Strickland:2011mw, Strickland:2011aa, Krouppa:2015yoa, Krouppa:2016jcl, Krouppa:2017jlg, Islam:2020gdv} as an effective model suitable for phenomenological analysis if the dissipative effects can be ignored.
The complex potential model works better than the U(1) case because of the aforementioned nature of the SU($N_c$) stochastic potential model: the dominance of octet states and the repulsive force between octet pair.

Another purpose is to show the symmetry properties of the SU($N_c$) stochastic potential model, such as global SU($N_c$), parity, and charge conjugation.
They are symmetries of the total system so that the evolution of the quarkonium density matrix respects these symmetries while each realization of the stochastic wave function does not.
Therefore, they are symmetries of the model which hold only after taking the event average.
In the case of $N_c=2$, we discover a peculiar ``event-by-event" symmetry due to the pseudo-reality of SU(2), whose transformation we call ``spinor conjugation".
Physically, this emergent symmetry results from the fact that the gluons cannot distinguish heavy quark and anti-quark because both of them have colors (up to some operation for the latter).
Therefore, the QGP environment is blind to an exchange of the color carriers of the system (heavy quark $\leftrightarrow$ heavy antiquark) and preserves an off-diagonal structure (``X") in the density matrix by the relative coordinates.

Our calculation is based on stochastic Schr\"odinger equation with SU($N_c$) noise with finite correlation length.
When the heavy quark pair is close to each other, one can make the dipole approximation in the Lindblad master equation equivalent to the stochastic potential model.
In the dipole limit, two groups have already published their results; the one is by directly solving the Lindblad equation \cite{Brambilla:2016wgg, Brambilla:2017zei} or by the quantum-jump method \cite{Brambilla:2020qwo,Brambilla:2021wkt, Omar:2021kra} and the other is by the stochastic Schr\"odinger equation \cite{Sharma:2019xum}\footnote{
However, we doubt if the stochastic potential description is available when the subspace (e.g. angular momentum and color) of the density matrix is projected.
See Appendix \ref{app:lindblad} for details.
}.
In principle, one has to check the consistency of the dipole approximation by monitoring the distribution of the relative distance between the pair.

This paper is organized as follows.
In Section \ref{sec:model}, we explain the basics of the SU($N_c$) stochastic potential model.
In Section \ref{sec:symmetries}, the symmetries of SU($N_c$) stochastic potential model is analyzed.
In Section \ref{sec:simulation}, we show the results of our one-dimensional simulation for the SU($N_c$) stochastic potential model in a static and an expanding background QGP.
In Section \ref{sec:conclusion}, we summarize the paper.
In the Appendix \ref{app:lindblad}, we provide a relation between the SU($N_c$) stochastic potential model and the Lindblad equation with an emphasis that the former is not available when the density matrix is projected to subspaces.
In the Appendix \ref{app:symmetries}, we give a summary of charge and spinor conjugations in the gauge theories and a relation of the latter to the $G$-parity.

\section{SU($N_c$) stochastic potential model} \label{sec:model}
In this section, we review the stochastic potential model including color degrees of freedom to describe quarkonium relative motion in the quark-gluon plasma \cite{Akamatsu:2014qsa}.
In Sec.~\ref{sec:simulation}, we show our numerical results for $N_c=3$ and $2$. 
We denote SU($N_c$) generator by $t^a$.
For SU(2) $t^a = \sigma^a/2$ ($a=1, 2, 3$) and for SU(3) $t^a = \lambda^a/2$ ($a=1, 2, \ldots, 8$), where $\sigma^a$ and $\lambda^a$ are Pauli and Gell-Mann matrices.

In the model, the Hamiltonian of quarkonium is given by
\begin{subequations}
\label{eq:hamiltonian}
\begin{align}
H_{\Theta}(\bm r, t) &\equiv  -\frac{{\bm\nabla}^2_{r}}{M}+V(\bm r)(t^a \otimes t^{a \ast})  +\Theta(\bm{r},t),\\
\Theta(\bm{r},t) &\equiv \theta^a(\bm{R}+\frac{\bm{r}}{2},t)(t^a\otimes 1)
-\theta^a(\bm{R}-\frac{\bm{r}}{2},t)(1\otimes t^{a*}),
\end{align}
\end{subequations}
where the noise with color index has properties
\begin{subequations}
\label{eq:sunnoiseproperty}
\begin{align}
&\langle\theta^a(\bm{x},t)\rangle=0, \\
&\langle\theta^a(\bm{x},t)\theta^b(\bm{x'},t')\rangle= D(\bm{x}-\bm{x'})\delta(t-t')\delta^{ab},
\end{align}
\end{subequations}
the spatial vectors $\bm{R}$ and $\bm{r}$ denote the center of mass and the relative coordinates of quarkonium, respectively, and $V(\bm r)$ denotes potential between the heavy quark and antiquark.
Since the noise correlations are translationally invariant, the $\bm R$-dependences in $H_{\Theta}$ and $\Theta$ are omitted here.
At the leading order in $g$ for the soft scale $r\sim 1/gT$, $V(\bm r)$ and $D(\bm r)$ are obtained as
\begin{subequations}
\label{eq:stoch_pot_pert}
\begin{align}
V(\bm r) &= -\frac{g^2}{4\pi r}e^{-m_D r}, \\
D(\bm r) &= g^2 T\int \frac{d^3k}{(2\pi)^3}\frac{\pi m_D^2 e^{i\bm k\cdot\bm r}}{k(k^2+m_D^2)^2},
\end{align}
\end{subequations}
with the Debye screening mass $m_D = gT\sqrt{N_c/3 + N_f/6}$ for $N_f$ massless flavors.
This shows that, at least in the weak coupling, the noise correlation length is of the order of the Debye screening length.

Using a tensor product basis for the color space, the wave function has $N_c\times N_c$ components $\Psi_{ij}$, where the first and second indices are in the $N_c$ and $N_c^*$ representations of SU($N_c$), respectively.
Physically, $i$ denotes heavy quark color and $j$ denotes heavy antiquark color.
The tensors in the Hamiltonian are explicitly given by
\begin{subequations}
\label{eq:tensor}
\begin{align}
(t^a \otimes t^{a \ast})_{ij,kl} &= (t^a )_{ik}(t^{a*})_{jl} = (t^a )_{ik}(t^{a})_{lj} \nonumber \\
&= \frac{1}{2} \left( \delta_{ij}\delta_{kl}- \frac{1}{N_c} \delta_{ik}\delta_{lj}\right), \\
(t^a \otimes 1)_{ij,kl} &= (t^a)_{ik}\delta_{jl}, \\
(1 \otimes t^{a \ast})_{ij,kl} &= \delta_{ik}(t^{a*})_{jl} = \delta_{ik}(t^{a})_{lj} .
\end{align}
\end{subequations}
The dynamics of quarkonium in the hot medium is governed by SU($N_c$) stochastic Schr\"odinger equation,
\begin{subequations}
\label{eq:sse3}
\begin{align}
i \frac{\partial}{\partial t}\Psi_{ij}(\bm{r},t) &= [H^{\rm eff}_{\Theta}(\bm r, t)]_{ij,kl} \Psi_{kl}(\bm{r},t), \\
H^{\rm eff}_{\Theta}(\bm r, t) &= -\frac{{\bm\nabla}^2_{r}}{M}-iC_{F} D(0) \\
& \quad +[V(\bm r)+iD(\bm{r})](t^a \otimes t^{a*})
+ \Theta(\bm r,t), \nonumber 
\end{align}
\end{subequations}
which is obtained by expanding the time evolution operator $e^{-i\Delta t H(\bm r, t)}$ up to the first order in $\Delta t$.
An equivalent representation for the stochastic Schr\"odinger equation is obtained by expressing the wave function as an $N_c\times N_c$ matrix $\Psi_{ij}=(\Psi)_{ij}$:
\begin{align}
\label{eq:sse4}
i \frac{\partial}{\partial t}\Psi(\bm{r},t) &=\left[ -\frac{{\bm\nabla}^2_{r}}{M}-iC_{F} D(0)\right]\Psi  \\
& \quad +\left[V(\bm r)+iD(\bm{r})\right] t^a \Psi t^a\nonumber \\
& \quad + \theta^a(\bm R + \frac{\bm r}{2},t) t^a \Psi
- \Psi t^a\theta^a(\bm R - \frac{\bm r}{2},t). \nonumber
\end{align}
The averaged wave function $\langle\Psi_{ij}\rangle$ evolves by non-hermitian Hamiltonian
\begin{subequations}
\begin{align}
i \frac{\partial}{\partial t}\langle\Psi_{ij}(\bm{r},t)\rangle &= \langle [H^{\rm eff}_{\Theta}(\bm r, t)]_{ij,kl}\rangle \langle\Psi_{kl}(\bm{r},t)\rangle, \\
\langle H^{\rm eff}_{\Theta}(\bm r, t)\rangle &= -\frac{{\bm\nabla}^2_{r}}{M}-iC_{F} D(0) \\
& \quad +[V(\bm r)+iD(\bm{r})](t^a \otimes t^{a*}), \nonumber 
\end{align}
\end{subequations}
from which the complex potential for the singlet is extracted as $C_F [V(\bm r) + i(D(\bm r)-D(0))]$.
The imaginary part of the potential for static heavy quark pair was first discovered in perturbative calculations at high temperatures \cite{Laine:2006ns, Beraudo:2007ky, Brambilla:2008cx} and then has been studied actively by lattice simulations \cite{Rothkopf:2011db, Burnier:2014ssa, Burnier:2016mxc, Petreczky:2018xuh, Bala:2019cqu} and in exotic setups such as in the presence of magnetic fields \cite{Bonati:2016kxj, Hasan:2017fmf, Singh:2017nfa}, etc.

\section{Symmetries of the Model}\label{sec:symmetries}
Here, we discuss symmetries of the SU($N_c$) stochastic potential model.
Namely, we show the model is invariant under global SU($N_c$), parity, and charge conjugation as it should because QCD is invariant under these transformations and because we consider environment with such symmetries, i.e. QGP with vanishing baryon chemical potential $\mu_B=0$ and without any external color and electromagnetic fields.
For $N_c=2$, the stochastic potential contains an additional non-trivial symmetry ``spinor conjugation" which derives from the pseudo-reality of SU(2) algebra.
What makes this symmetry special is that it holds in an event-by-event manner, so that it places a stronger constraint than the normal symmetries.
Schematically, such an event-by-event symmetry introduces constraints between
\begin{subequations}
\label{eq:sym_constraint_ebe}
\begin{align}
\rho(\bm r, \bm r', t) &=\langle \Psi(\bm r,t)\Psi(\bm r',t)^*\rangle, \\
\rho^{\mathcal U}(\bm r, \bm r', t)&= \langle [\mathcal U\Psi(\bm r,t)]\Psi(\bm r',t)^*\rangle, \\
\rho^{\mathcal U^*}(\bm r, \bm r', t)&= \langle \Psi(\bm r,t)[\mathcal U\Psi(\bm r',t)]^*\rangle,
\end{align}
\end{subequations}
in addition to the usual one
\begin{align}
\label{eq:sym_constraint_usual}
\rho^{\mathcal {UU}^*}(\bm r, \bm r', t)= \langle [\mathcal U\Psi(\bm r,t)] [\mathcal U\Psi(\bm r',t)]^*\rangle,
\end{align}
where $\mathcal U$ is the symmetry transformation.
What distinguishes the usual symmetry and the event-by-event symmetry is that the former involves transformation on the environment as well while the latter leaves the environment unchanged.
Note that since the constraints between \eqref{eq:sym_constraint_ebe} and \eqref{eq:sym_constraint_usual} follow from symmetry properties, they hold beyond the decoherence dynamics, i.e., when quantum dissipative effects are included.

\subsection{Global SU($N_c$) symmetry}\label{sec:symmetry_suN}
First, we show that the stochastic potential model has global SU($N_c$) symmetry.
Under the transformation, the wave function transforms as
\begin{align}
\psi_{ij}^U(\bm r,t) = U_{ik}\psi_{kl}(\bm r,t)U^{\dagger}_{lj}, \quad
U \in {\rm SU}(N_c).
\end{align}
Since the noise field originates from $A_0$ field, it also transforms as
\begin{subequations}
\begin{align}
\theta^{Ua}(\bm x,t)t^a
&=U\theta^{a}(\bm x,t)t^aU^{\dagger}, \\
\Theta^U(\bm r, t) &= \theta^{Ua}(\bm{R}+\frac{\bm{r}}{2},t)(t^{a}\otimes 1) \nonumber \\
& \quad -\theta^{Ua}(\bm{R}-\frac{\bm{r}}{2},t)(1\otimes t^{a*}),
\end{align}
\end{subequations}
where the explicit form of the former is
\begin{align}
\theta^{Ua}(\bm r,t) = 2{\rm tr}\left[
t^a U\theta^b(\bm r,t)t^bU^{\dagger}
\right]
=K_{ab}\theta^b(\bm r,t).
\end{align}
Note that $K_{ab}$ is an orthogonal matrix\footnote{
Here is the proof:
\begin{subequations}
\begin{align}
&K_{ab}^*=2{\rm tr}\left[t^a U t^b U^{\dagger}\right]^{\dagger} = 2{\rm tr}\left[U t^b U^{\dagger} t^a\right]=K_{ab},\\
&K_{ab}K_{cb} = 4(t^a)_{ij}U_{jk}(t^b)_{kl}U^{\dagger}_{li} 
(t^c)_{mn}U_{np}(t^b)_{pq}U^{\dagger}_{qm}  \\
& \quad = 4(t^a)_{ij}U_{jk}U^{\dagger}_{li}(t^c)_{mn}U_{np}U^{\dagger}_{qm}
\frac{1}{2}\left(\delta_{kq}\delta_{lp} - \frac{1}{N_c}\delta_{kl}\delta_{pq}\right) =\delta_{ab}. \nonumber
\end{align}
\end{subequations}
}
so that $\theta^{Ua}$ is real and satisfies
\begin{align}
\langle\theta^{Ua}(\bm x,t)\theta^{Ub}(\bm x', t')\rangle
=D(\bm x-\bm x')\delta(t-t')\delta^{ab}.
\end{align}
Under the global SU($N_c$) transformation, the Hamiltonian \eqref{eq:hamiltonian} becomes
\begin{align}
H^U_{\Theta} = -\frac{{\bm\nabla}^2_{r}}{M}+V(\bm r)(t^a \otimes t^{a \ast})  +\Theta^U(\bm{r},t).
\end{align}
and is invariant in the statistical sense while it is not on the event-by-event basis, i.e. for a given noise field, because
\begin{align}
\Theta(\bm{r},t)\neq \Theta^U(\bm{r},t).
\end{align}
Therefore, the global SU($N_c$) symmetry constrains the density matrix\footnote{
Since $U$ is not an event-by-event symmetry, we use a notation $\rho^{U}$ to mean $\rho^{UU^*}$ in the above.
The same applies for parity and charge conjugation.
}
\begin{align}
\label{eq:dmat_color}
\rho^U_{ij,kl}(\bm r, \bm r', t)
&=U_{im}U^{\dagger}_{nj}\rho_{mn,pq}(\bm r, \bm r', t)U^{\dagger}_{pk}U_{lq}\nonumber \\
&=\rho_{ij,kl}(\bm r, \bm r', t),
\end{align}
if the initial density matrix satisfies the same relation.
Noting that the invariant tensors of rank $(2,2)$ are $\delta_{ij}\delta_{kl}$ and $\delta_{ik}\delta_{jl}$, the density matrix is always decomposed by their linear combination if it is initially.
For example, if the initial density matrix is
\begin{align}
\hspace{-1mm}
\rho_{ij,kl}(\bm r, \bm r', 0)
= \rho_s(\bm r, \bm r', 0) P^{(s)}_{ij,kl} + \rho_o(\bm r, \bm r', 0) P^{(o)}_{ij,kl},
\end{align}
with singlet and octet (or ($N_c^2-1$)-multiplet) projectors $P^{(s)}_{ij,kl} = \frac{\delta_{ij}\delta_{kl}}{N_c}$ and $P^{(o)}_{ij,kl} = \delta_{ik}\delta_{jl} - \frac{\delta_{ij}\delta_{kl}}{N_c}$, it is invariant under the global SU($N_c$) transformation.
Then the density matrix at later times is also given by two density matrices $\rho_{s/o}(\bm r,\bm r', t)$:
\begin{align}
\label{eq:dmat_tensors}
\rho_{ij,kl}(\bm r, \bm r', t)
= \rho_s(\bm r, \bm r', t) P^{(s)}_{ij,kl} + \rho_o(\bm r, \bm r', t) P^{(o)}_{ij,kl}.
\end{align}

Before closing, let us briefly mention the properties under a time-dependent ${\rm SU}(N_c)$ transformation, i.e., a spatially uniform ${\rm SU}(N_c)$ gauge transformation.
The wave function transforms as
\begin{align}
\psi_{ij}^G(\bm r,t) = U_{ik}(t)\psi_{kl}(\bm r,t)U^{\dagger}_{lj}(t), \quad
U(t) \in {\rm SU}(N_c).
\end{align}
while the noise field transforms as 
\begin{align}
\theta^{Ga}(\bm x,t) t^a &= U(t)\left[\theta^a(\bm x,t)t^a-i\partial_t\right]U^{\dagger}(t)\nonumber \\
&\equiv \theta'^{Ga}(\bm x,t) t^a + i \dot U(t) U^{\dagger}(t) .
\end{align}
The latter follows from the fact that the noise field models the fluctuations of $A_0$ field.
Note that the statistical property of $\theta'^{Ga}$ is identical to that of $\theta^a$.
The Hamiltonian is then written as
\begin{subequations}
\begin{align}
H_{\Theta}^G &= -\frac{{\bm\nabla}^2_{r}}{M}+V(\bm r)(t^a \otimes t^{a \ast})  +\Theta'^G(\bm{r},t) \\
& \quad   +[i\dot U(t) U^{\dagger}(t)]\otimes 1 -1\otimes [i\dot U(t)U^{\dagger}(t)]^* \nonumber,\\
\Theta'^G(\bm{r}, t) &= \theta'^{Ga}(\bm{R}+\frac{\bm{r}}{2},t)(t^{a}\otimes 1) \nonumber \\
& \quad -\theta'^{Ga}(\bm{R}-\frac{\bm{r}}{2},t)(1\otimes t^{a*}),
\end{align}
\end{subequations}
and the time-evolution equation for $\psi_{ij}^G(\bm r,t)$ is invariant in the statistical sense.
It means that there is a freedom to arbitrarily rotate the colors of $\psi_{ij}(\bm r,t)$ at each time without changing the physical contents.
Therefore, from the density matrix of quarkonia, one should compute observables defined with
\begin{subequations}
\begin{align}
\rho_s(\bm r, \bm r', t) \equiv \rho_{ij,kl}(\bm r, \bm r', t)P^{(s)}_{ij,kl},\\
\rho_o(\bm r, \bm r', t) \equiv \rho_{ij,kl}(\bm r, \bm r', t)P^{(o)}_{ij,kl},
\end{align}
\end{subequations}
which are invariant under this class of gauge transformations.

\subsection{Parity}\label{sec:symmetry_P}
Here let us consider the parity transformation (around $\bm R$).
Under parity, the wave function transforms as:
\begin{align}
\Psi_{ij}^P(\bm r, t) = -\Psi_{ij}(-\bm r,t),
\end{align}
where $-1$ accounts for relative parity between heavy quark and antiquark.
In addition, the noise field transforms as
\begin{align}
\theta^{Pa}(\bm x,t) = \theta^{a}(2\bm R -\bm x,t), \quad
\Theta^P(\bm r,t) = \Theta(-\bm r,t).
\end{align}
Then the Hamiltonian \eqref{eq:hamiltonian} becomes
\begin{align}
H^P_{\Theta} = -\frac{{\bm\nabla}^2_{r}}{M}+V(\bm r)(t^a \otimes t^{a \ast})  +\Theta^P(\bm{r},t).
\end{align}
Since the statistical average of $\theta^{Pa}$ obeys
\begin{align}
\langle\theta^{Pa}(\bm x,t)\theta^{Pb}(\bm x',t')\rangle
&=D(\bm x'-\bm x)\delta(t'-t)\delta^{ab} \nonumber \\
&=D(\bm x-\bm x')\delta(t-t')\delta^{ab},
\end{align}
the Hamiltonian \eqref{eq:hamiltonian} is parity invariant in the statistical sense.
However, parity invariance does not hold on the event-by-event basis, because
\begin{align}
\Theta^P(\bm r,t) \neq \Theta(\bm r,t).
\end{align}
Therefore, the parity invariance only constrains the density matrix: if the initial density matrix is parity invariant
\begin{align}
\rho_{ij,kl}^P(\bm r, \bm r', 0) = \rho_{ij,kl}(-\bm r, -\bm r', 0)= \rho_{ij,kl}(\bm r, \bm r', 0),
\end{align}
then it remains so
\label{eq:dmat_parity}
\begin{align}
\rho_{ij,kl}^P(\bm r, \bm r', t) =\rho_{ij,kl}(-\bm r, -\bm r', t) =\rho_{ij,kl}(\bm r, \bm r', t) .
\end{align}

\subsection{Charge conjugation}\label{sec:symmetry_C}
Next, let us describe the charge conjugation.
Under the charge conjugation, a heavy quark with color $i$ (e.g. red) is transformed into a heavy antiquark with anti-color $i$ (e.g. anti-red).
Therefore, the wave function transforms as (see Eq.~\eqref{eq:Psi_C})
\begin{align}
\Psi_{ij}^C(\bm r,t) = -\Psi_{ji}(-\bm r,t),
\end{align}
where $-1$ takes care of the interchange of fermions.
The noise field also transforms as
\begin{subequations}
\begin{align}
\theta^{Ca}(\bm x,t)t^a &= - \theta^a(\bm x,t)t^{a*}, \\
\Theta^C(\bm r, t) &= \theta^{Ca}(\bm{R}+\frac{\bm{r}}{2},t)(t^{a}\otimes 1) \nonumber \\
& \quad -\theta^{Ca}(\bm{R}-\frac{\bm{r}}{2},t)(1\otimes t^{a*}),
\end{align}
\end{subequations}
where the former is written explicitly as
\begin{align}
\theta^{Ca}(\bm r,t) = - 2{\rm tr}\left[t^a\theta^b(\bm r,t)t^{b*}\right] = M_{ab}\theta^b(\bm r,t).
\end{align}
Note that $M_{ab}$ is an orthogonal and symmetric matrix\footnote{
It is trivial to show $M_{ab}^* = M_{ba} = M_{ab}$ while the orthogonality follows from
\begin{align}
M_{ab}M_{cb} &= 4(t^a)_{ij}(t^b)_{ij}(t^c)_{kl}(t^b)_{kl} \nonumber \\
&=4(t^a)_{ij}(t^c)_{kl} \frac{1}{2} \left( \delta_{il}\delta_{jk}- \frac{1}{N_c} \delta_{ij}\delta_{kl}\right)
=\delta_{ab}.
\end{align}
} so that $\theta^{Ca}$ is real and satisfies
\begin{align}
\langle\theta^{Ca}(\bm x,t)\theta^{Cb}(\bm x',t')\rangle
&=D(\bm x-\bm x')\delta(t-t')\delta^{ab}.
\end{align}
Then the Hamiltonian \eqref{eq:hamiltonian} is transformed into
\begin{align}
H^C_{\Theta} &=-\frac{{\bm\nabla}^2_{r}}{M}+V(\bm r)(t^{a*} \otimes t^a)  \nonumber \\
& \quad +\theta^{a}(\bm{R}-\frac{\bm{r}}{2},t)(1\otimes t^{a}) 
-\theta^{a}(\bm{R}+\frac{\bm{r}}{2},t)(t^{a*}\otimes 1) \nonumber\\
&=-\frac{{\bm\nabla}^2_{r}}{M}+V(\bm r)(t^{a} \otimes t^{a*}) + \Theta^C(\bm r,t),
\end{align}
and is shown to be invariant under the charge conjugation in the statistical sense.
Note here again that the charge conjugation is not an event-by-event symmetry, because
\begin{align}
\Theta^C(\bm r,t) \neq  \Theta(\bm r,t).
\end{align}
It only constrains the density matrix by
\begin{align}
\label{eq:dmat_charge}
\rho_{ij,kl}^C(\bm r, \bm r', t) =\rho_{ji,lk}(-\bm r, -\bm r', t) =\rho_{ij,kl}(\bm r, \bm r', t) .
\end{align}
if the latter equality holds at $t=0$.

Let us finally comment on the the charge conjugation symmetry of quarkonium stochastic potential in a QGP with finite baryon chemical potential.
Since the coefficients in the master equation for quarkonium is defined by $C$-even quantities, i.e. two-point functions of gluons, the quarkonium is influenced only by $C$-even properties of the QGP.
Therefore, the Hamiltonian \eqref{eq:hamiltonian} has charge conjugation symmetry even for $\mu_B\neq 0$.

\subsection{``Spinor conjugation" for $N_c=2$}\label{sec:symmetry_S}
The global SU($N_c$), parity, and charge conjugation are symmetries of the Hamiltonian \eqref{eq:hamiltonian} in the statistical sense.
The reason they do not hold on the event-by-event basis is that it involves a transformation of the noise field.
Here we find a transformation which can deceive the gluons and thus leaves the noise field unaltered.

Using the pseudo-reality of SU(2), wave function for a heavy quark with {\it anti-color} $j$ at $\bm R - \bm r/2$ and a heavy antiquark with {\it color} $i$ at $\bm R + \bm r/2$ can be obtained by (see Eq.~\eqref{eq:Psi_S})
\begin{align}
\label{eq:spinorconj_tensor}
-\epsilon_{il}\epsilon_{jk}\Psi_{kl}(-\bm r,t) \equiv \Psi^{S}_{ij}(\bm r,t).
\end{align}
In the matrix notation, this transformation is
\begin{align}
\label{eq:spinorconj_matrix}
\Psi^{S}(\bm r,t) &\equiv -\sigma^2 [\Psi(-\bm r, t)]^T \sigma^2, \quad
\sigma^2 \equiv \begin{pmatrix}
0 & -i \\
i & 0
\end{pmatrix},
\end{align}
where $T$ denotes the transposition.
Since $\Psi_{ij}(\bm r,t)$ is a wave function for a heavy quark with {\it color} $i$ at $\bm R + \bm r/2$ and a heavy antiquark with {\it anti-color} $j$ at $\bm R - \bm r/2$, this transformation only exchanges the carriers of the (anti)colors.
This is the reason why we call this transformation ``spinor conjugation".
The carriers of the colors are not distinguished by the gluons so that spinor conjugation is a symmetry in any given gluon backgrounds.

We can show the invariance of the stochastic Schr\"odinger equations \eqref{eq:sse3} and \eqref{eq:sse4} under the spinor conjugation.
Since the matrix form \eqref{eq:sse4} is easier to handle, we only show the proof for this case.
By substituting
\begin{align}
\label{eq:spinorconj_inv}
\Psi(\bm r,t) = -\sigma^2 [\Psi^{S}(-\bm r, t)]^T\sigma^2,
\end{align}
we get
\begin{align}
i \frac{\partial}{\partial t}\sigma^2[\Psi^{S}(\bm{r},t)]^T\sigma^2 
& =\left[ -\frac{{\bm\nabla}^2_{r}}{M}-iC_{F} D(0)\right]\sigma^2[\Psi^{S}]^T\sigma^2  \nonumber \\
& \hspace{-5mm}\quad +\left[V(\bm r)+iD(\bm{r})\right] t^a\sigma^2 [\Psi^{S}]^T\sigma^2 t^a\nonumber \\
&\hspace{-5mm} \quad + \theta^a(\bm R - \frac{\bm r}{2},t) t^a\sigma^2 [\Psi^{S}]^T\sigma^2\nonumber \\
& \hspace{-5mm}\quad - \sigma^2[\Psi^{S}]^T \sigma^2 t^a\theta^a(\bm R + \frac{\bm r}{2},t). 
\end{align}
Using the property
\begin{align}
t^a \sigma^2 = -\sigma^2(t^a)^T, \quad
\sigma^2 t^a = - (t^a)^T\sigma^2,
\end{align}
we obtain
\begin{align}
i \frac{\partial}{\partial t}\Psi^{S}(\bm{r},t) &=\left[ -\frac{{\bm\nabla}^2_{r}}{M}-iC_{F} D(0)\right]\Psi^{S}  \\
& \hspace{-5mm}\quad +\left[V(\bm r)+iD(\bm{r})\right] t^a \Psi^{S} t^a\nonumber \\
& \hspace{-5mm}\quad + \theta^a(\bm R + \frac{\bm r}{2},t) t^a \Psi^{S}
- \Psi^{S} t^a\theta^a(\bm R - \frac{\bm r}{2},t), \nonumber
\end{align}
which proves that the spinor conjugation is an event-by-event symmetry.
As is clear from Eqs.~\eqref{eq:spinorconj_matrix} and \eqref{eq:spinorconj_inv}, the eigenvalues of the spinor conjugation are $\pm 1$.
An explicit tensor analysis of Eq.~\eqref{eq:spinorconj_tensor} shows that $S$-even states are parity-odd color-singlet and parity-even color-triplet states, and $S$-odd states are parity-even color-singlet and parity-odd color-triplet states.
Such states remain $S$-even/odd throughout the time evolution.

Since the spinor conjugation is an event-by-event symmetry, we can obtain constraints from three transformations on the density matrix:
\begin{subequations}
\label{eq:dmat_spinor}
\begin{align}
\rho_{ij,kl}^{S}(\bm r, \bm r', t) &= - \epsilon_{in}\epsilon_{jm}\rho_{mn,kl}(-\bm r, \bm r', t), \\
\rho_{ij,kl}^{S^*}(\bm r, \bm r', t) &= - \rho_{ij,mn}(\bm r, -\bm r', t) \epsilon_{kn}\epsilon_{lm}, \\
\rho_{ij,kl}^{SS^*}(\bm r, \bm r', t) &= \epsilon_{in}\epsilon_{jm}\rho_{mn,pq}(-\bm r, -\bm r', t)\epsilon_{kq}\epsilon_{lp}.
\end{align}
\end{subequations}
The first two take spinor conjugation for either the wave function or its complex conjugate and the last takes for the both.
The spinor conjugation symmetry requires
\begin{subequations}
\label{eq:dmat_spinor_even}
\begin{align}
\rho_{ij,kl}^{S}(\bm r, \bm r', t) &=\rho_{ij,kl}^{S^*}(\bm r, \bm r', t) =  \rho_{ij,kl}(\bm r, \bm r', t), \\
\rho_{ij,kl}^{SS^*}(\bm r, \bm r', t)&= \rho_{ij,kl}(\bm r, \bm r', t),
\end{align}
\end{subequations}
if the initial density matrix is composed only of $S$-even states, and
\begin{subequations}
\label{eq:dmat_spinor_odd}
\begin{align}
\rho_{ij,kl}^{S}(\bm r, \bm r', t) &=\rho_{ij,kl}^{S^*}(\bm r, \bm r', t) = - \rho_{ij,kl}(\bm r, \bm r', t), \\
\rho_{ij,kl}^{SS^*}(\bm r, \bm r', t)&= \rho_{ij,kl}(\bm r, \bm r', t),
\end{align}
\end{subequations}
if the initial density matrix is composed only of $S$-odd states. 
By combining Eq.~\eqref{eq:dmat_tensors} for $N_c=2$ and Eqs.~\eqref{eq:dmat_spinor_even} and \eqref{eq:dmat_spinor_odd}, one can show that for $S$-even/odd initial states\footnote{
We write $\rho_{t}(\bm r,\bm r',t )$ and $P^{(t)}_{ij,kl}$ instead of $\rho_{o}(\bm r,\bm r',t )$ and $P^{(o)}_{ij,kl}$.
}
\begin{subequations}
\label{eq:dmat_spinor_st}
\begin{align}
\rho_s(\bm r,\bm r',t ) &= \mp\rho_s(-\bm r,\bm r',t ) = \mp\rho_s(\bm r, -\bm r',t ), \\
\rho_t(\bm r,\bm r',t ) &= \pm\rho_t(-\bm r,\bm r',t ) = \pm\rho_t(\bm r, -\bm r',t ),
\end{align}
\end{subequations}
which follows from
\begin{subequations}
\begin{align}
-\epsilon_{in}\epsilon_{jm}P^{(s)}_{mn,kl} &= -P^{(s)}_{ij,kl}, \\
-\epsilon_{in}\epsilon_{jm}P^{(t)}_{mn,kl} &= P^{(t)}_{ij,kl}.
\end{align}
\end{subequations}
An interesting observation is that the spinor conjugation insures finite off-diagonal elements of density matrix even after decoherence proceeds.
Structure of the symmetry protected off-diagonal element is explicitly confirmed by the numerical simulation in the next section.

So far, we have ignored spin degrees of freedom in charge and spinor conjugations.
Indeed, we can show that they are ``spin-less" versions of the fundamental symmetries.
In the Appendix \ref{app:symmetries}, we summarize the charge and spinor conjugations of the gauge theory and make a brief comparison to the $G$-parity.

\section {Numerical simulation} \label{sec:simulation}
We solve the color SU(3) stochastic Schr\"odinger equation \eqref{eq:sse3} numerically in one spatial dimension.
The numerical setup for bottomonium (charmonium) is as follows:
system of $-2.56 {\rm fm} \leq x \leq 2.56 {\rm fm}$ ($-5.12 {\rm fm} \leq x \leq 5.12 {\rm fm}$) is discretized with 512 cells (1024 cells) of size $\Delta x = 0.01$ fm with the periodic boundary condition, and the wave functions are updated from $t=0$ to $t=9$ fm by 90000 steps of $\Delta t =0.0001$ fm.
The spatial size of 5.12 fm (10.24 fm) is large enough to accommodate bound state wave functions in our computations.
We collect 1000 wave functions evolved by the stochastic Schr\"odinger equation and take their average to calculate the thermal ensemble average.
As in our previous paper \cite{Kajimoto:2017rel}, we parametrize two functions $V(x)$ and $D(x)$ of the model by
\begin{subequations}
\label{eq:stoch_pot_model}
\begin{align}
V(x) &= -\frac{\alpha_{\rm eff}}{|x|}\exp\left(-m_D |x|\right), \quad \\
D(x) &= \gamma \exp\left(-|x|^2/l_{\rm corr}^2\right),
\end{align}
\end{subequations}
motivated by perturbative results \eqref{eq:stoch_pot_pert}\footnote{
The Coulomb singularity of $V(x)$ is regulated by replacing $|x|$ with $\sqrt{x^2+1/M^2}$ in $V(x)$, where $M$ is the heavy quark mass.
}.
Perturbative estimate for these parameters is
\begin{subequations}
\begin{align}
\alpha_{\rm eff} &= \frac{g^2}{4\pi}, \quad
m_D = gT\sqrt{\frac{N_c}{3} + \frac{N_f}{6}}, \\
\gamma &= D(0) = \frac{g^2 T}{4\pi}, \quad
l_{\rm corr} \sim 1/m_D.
\end{align}
\end{subequations}
We thus assume the relation $\gamma = \alpha_{\rm eff}T$ and $l_{\rm corr}=1/m_D$ with $\alpha_{\rm eff} = 0.3$ and $m_D=T$ as a reasonable choice (Table \ref{table:parameters}).
At the same time, we loose the relation $l_{\rm corr} =1/m_D$ in order to see how the decoherence depends on $l_{\rm corr}$.

In Sec.~\ref{sec:simulation_static}, we show the numerical results for bottomonium evolution in a static QGP with $T=0.4$ GeV.
In Sec.~\ref{sec:simulation_bjorken}, we show the results for bottomonium and charmonium in a Bjorken expanding QGP whose temperature decreases in time as
\begin{align}
T(t)=T_0 \left(\frac{t_0}{t_0+t}\right)^{1/3},
\label{eq:T}
\end{align}
with the initial temperature being $T_0=0.4$ GeV and the initial time $t_0= 1$ fm.
In Sec.~\ref{sec:simulation_nandrho}, we demonstrate the spatial distribution and density matrix for bottomonium in the Bjorken expansion.

\begin{table}
\caption{Parameters in the model.}
\label{table:parameters}
\begin{center}
\begin{tabular}{cccc|cc} \hline \hline
\ ${\alpha}_{\rm eff}$ \	& \ $m_{D}$ \	& \ $\gamma$ \	& \ $l_{\rm corr}$ \  & $M_b$   & $M_c$  \\ \hline
$0.3$ 				& $T$ 		& $0.3T$ 		& $1/T$ & \ $4.8$ GeV \ & \ $1.18$ GeV \	 \\ \hline
\end{tabular}
\end{center}
\end{table}

\subsection{Quarkonium in a static QGP} \label{sec:simulation_static}
We compute the time evolution of bottomonium wave functions by the stochastic Schr\"odinger equation \eqref{eq:sse3}.
The bottom quark mass is $M_b=4.8$ GeV and the constant temperature is $T=0.4$ GeV.
We change the noise correlation length $l_{\rm corr} = 0.04, 0.16, 0.32, 0.48, 0.96$ fm to study how the results depend on $l_{\rm corr}$.
If we take the default value $l_{\rm corr} = 1/m_D$, $l_{\rm corr}$ is about $0.5$ fm.
For the other parameters, we adopt the values in the Table \ref{table:parameters} with $T=0.4$ GeV.

\begin{figure}[t]
\centering
\includegraphics[clip,width=8.5cm,bb= 0 0 360 252]{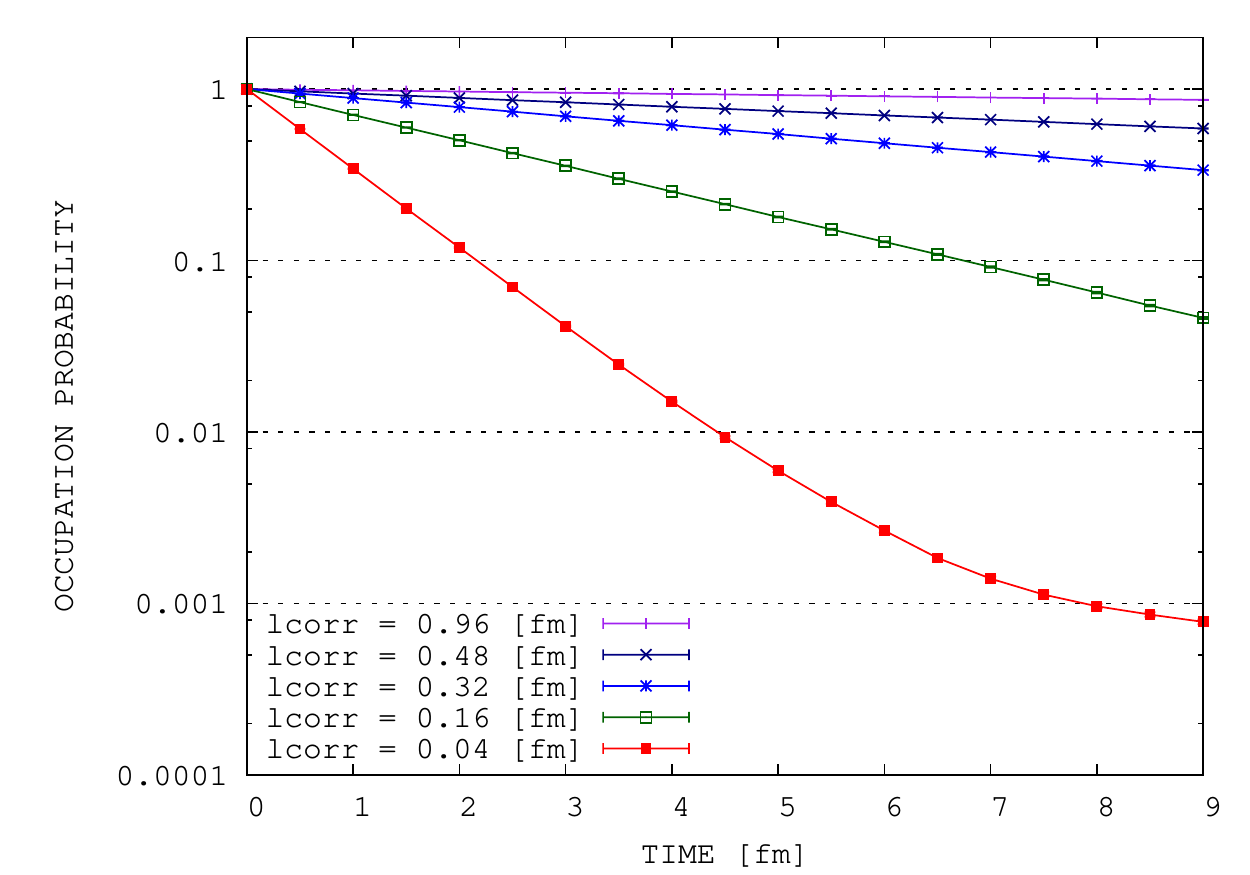}
\caption{
Time evolution of the ground state probability of bottomonium.
The initial and projected state is the ground state in the Debye screened potential at $T=0.4$ GeV.
The noise correlation length $l_{\rm corr}$ is varied from 0.04 fm to 0.96 fm.
}
\label{fig:ftsu3}
\end{figure}

In Fig. \ref{fig:ftsu3}, we show the time evolution of the occupation of color-singlet ground state in the Debye screened potential $V(x)$.
The simulation starts from the color-singlet ground state.
Note that with this setup, the occupation probability changes from unity entirely due to the noise term in the stochastic Schr\"odinger equation.
We change the noise correlation length $l_{\rm corr}$ to see how the decoherence is affected.
Since the ground state wave function is extended over $r_{\bar b b}\simeq 0.20$ fm, the decoherence by the noise becomes ineffective for $l_{\rm corr} = 0.32, 0.48, 0.96$ fm.
In other words, $l_{\rm corr}$ sets a color resolution scale of the QGP environment.

\begin{figure}
\begin{center}
\includegraphics[clip,width=8.5cm,bb= 0 0 355 245]{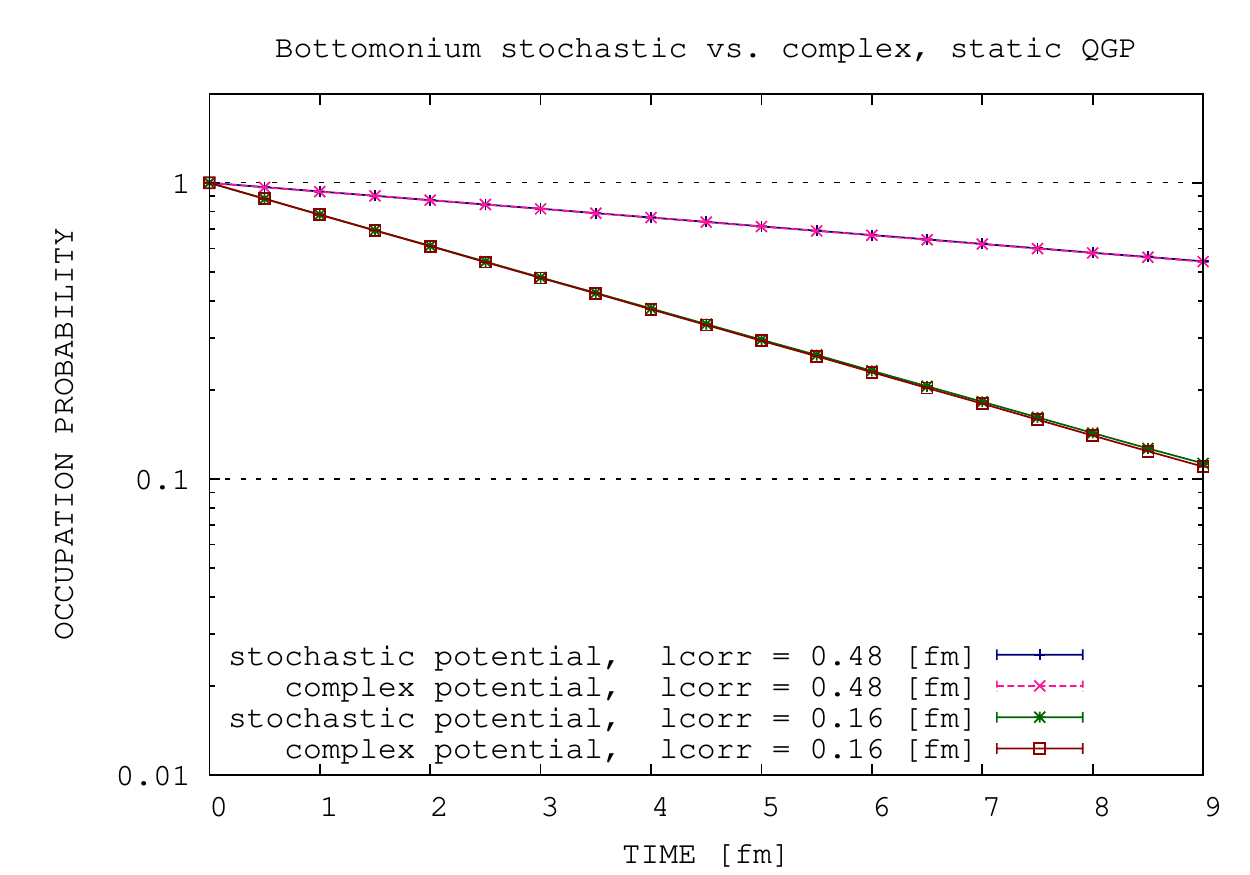}
\caption{
Time evolution of the occupation probability of the bottomonium ground state, with the noise correlation length $l_{\rm corr} = 0.48$ and $0.16$ fm and the temperature $T=0.4$ GeV.
Solid lines are the probabilities calculated in the stochastic potential model and dotted lines are those by the Schr\"odinger equation with the complex potential.
}
\label{fig:ft3com}
\end{center}
\end{figure}

In Fig. \ref{fig:ft3com}, we compare time evolution of the occupation probability for the stochastic potential and the complex potential with the parameter $l_{\rm corr}=0.48$ and $0.16$ fm.
While it is no doubt that the stochastic potential model is a more faithful description of quarkonium in the QGP, the agreement indicates that as long as we are to obtain the survival probability, we can use the complex potential instead of the stochastic potential.
The agreement for the SU(3) stochastic potential is better than our previous result with U(1) stochastic potential \cite{ Kajimoto:2017rel}.
The reason is as follows.
It takes at least two scatterings for the color-singlet ground state to get once excited to a color-octet and then get de-excited back to the color-singlet state.
The necessity of two scatterings is shared with the U(1) case.
For the SU(3) case, however, only a fraction ($1/(N_c^2-1)=1/8$ at maximum) of the second scattering turns the color-octet into the color-singlet\footnote{
In the SU($N_c$) stochastic potential model for a heavy quark pair at rest ($M\to\infty$) with relative distance $r$ fixed, the fraction $f_{o\to s}(r)$ that an octet turns into the singlet in one scattering is computed as
\begin{align}
\label{eq:fraction_os}
f_{o\to s} (r)=\frac{\langle{\rm tr}P^{(s)}\Theta(r)P^{(o)}\Theta(r)\rangle}{\langle{\rm tr}\Theta(r)P^{(o)}\Theta(r)\rangle} = \frac{D(0) - D(r)}{(N_c^2-1)D(0) + D(r)}.
\end{align}
Since $D(r)$ is a monotonically decreasing positive function for $r\geq 0$, the minimum and maximum of $f_{o\to s}(r)$ are $f_{o\to s}(0) = 0$ and $f_{o\to s}(\infty) = 1/(N_c^2-1)$.
The suppression $f_{o\to s} \sim 1/N_c^2$ is supported beyond the weak coupling regime by the large $N_c$ argument \cite{Escobedo:2020tuc}.
}.
Therefore, the time interval between the effective two scatterings, which bring back to the color singlet state, becomes longer for the SU(3) case.
Furthermore, since the potential between the color-octet heavy quark pair is repulsive and the pair gets far apart.
When the heavy quark pair turns into the color-singlet state, the wave function is already extended and its overlap to the localized ground state becomes very small.
In short, when $N_c$ is large, the color-singlet ground state almost never returns back to the original state once it is excited to the octet (or $N_c^2-1$) state.
This is precisely where the complex potential is applicable so that better agreement is expected between the stochastic potential and the complex potential.
This logic is applicable to the calculation of survival probability of any color-singlet bound states, in principle.
However, if one needs to calculate the transition from the initial state, the complex potential does not work so well, as is expected.
In Sec.~\ref{sec:simulation_bjorken}, we demonstrate such differences.

\begin{figure}
\centering
\includegraphics[clip,width=8.5cm,bb= 0 0 440 320]{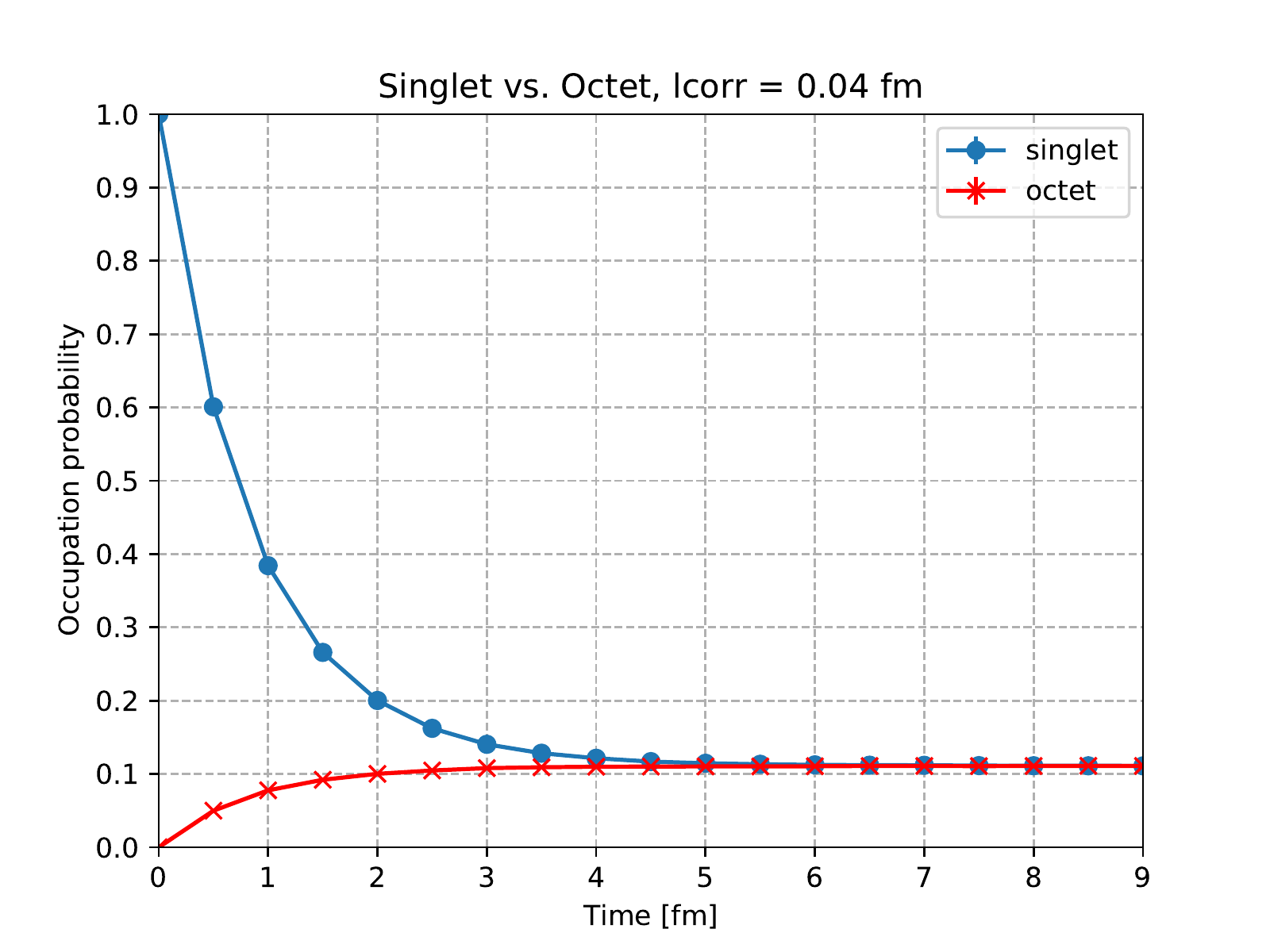}
\caption{
Time evolution of the occupation number of color-singlet and color-octet states for $l_{\rm corr}=0.04$ fm.
The color-octet occupation is the average value for the 8 color-octet states.
}
\label{fig:ft3_so}
\end{figure}

In Fig. \ref{fig:ft3_so}, we show the time evolution of color state occupation for the smallest $l_{\rm corr}$ in our simulation, namely $l_{\rm corr}=0.04$ fm.
The initial condition is again the color-singlet ground state.
Since we choose the smallest $l_{\rm corr}$, the decoherence proceeds fast enough to see the color space randomization in about 5 fm.
Note that the occupation probability of the color-octet is the average value for the 8 states\footnote{
Before taking average, we confirm that the 8 states are almost degenerated and indistinguishable.
}
so that each color state occupation probability approaches $1/9$.

\subsection{Quarkonium in a Bjorken-expanding QGP} \label{sec:simulation_bjorken}
Here we compute the time evolution of a bottomonium and a charmonium in a Bjorken expanding medium.
The bottom and charm quark masses are $M_b=4.8$ GeV and $M_c=1.18$ GeV.
We assume that the temperature decreases in time as
\begin{align}
T(t)=T_0 \left(\frac{t_0}{t_0+t}\right)^{1/3}, \quad
T_0 = 0.4 \ {\rm GeV}, \quad
t_0 = 1 \ {\rm fm}.
\end{align}
The parameters of the stochastic Schr\"odinger equation \eqref{eq:sse3} change accordingly.
In this simulation, we prepare the initial wave functions as the eigenstates of the (regularized) vacuum Cornell potential for the color singlet
\footnote{
The Coulomb singularity of $V_{\rm vac}(x)$ is regulated by replacing $|x|$ with $\sqrt{x^2+1/M^2}$ in $V_{\rm vac}(x)$, where $M$ is the heavy quark mass.
Note that  $t^a\otimes t^{a*}=C_F$ for the color singlet.
}:
\begin{align}
V_{\rm vac}(x) = -\frac{C_F\alpha_{\rm eff}}{|x|} + \sigma |x|, \quad
\sigma = 1 \ {\rm GeV}/{\rm fm},
\end{align}
with $\alpha_{\rm eff}=0.3$ as in the Table \ref{table:parameters}.
We also compute the occupation probability of these eigenstates, not of the eigenstates of the in-medium potential.
It is still not clear theoretically what initial wave function should be used.
Here we take the vacuum bound states for simplicity, but some take an octet wave packet as an initial condition.

\begin{figure}
\begin{center}
\begin{tabular}{c}
\includegraphics[clip,width=9.0cm,bb= 0 0 380 270]{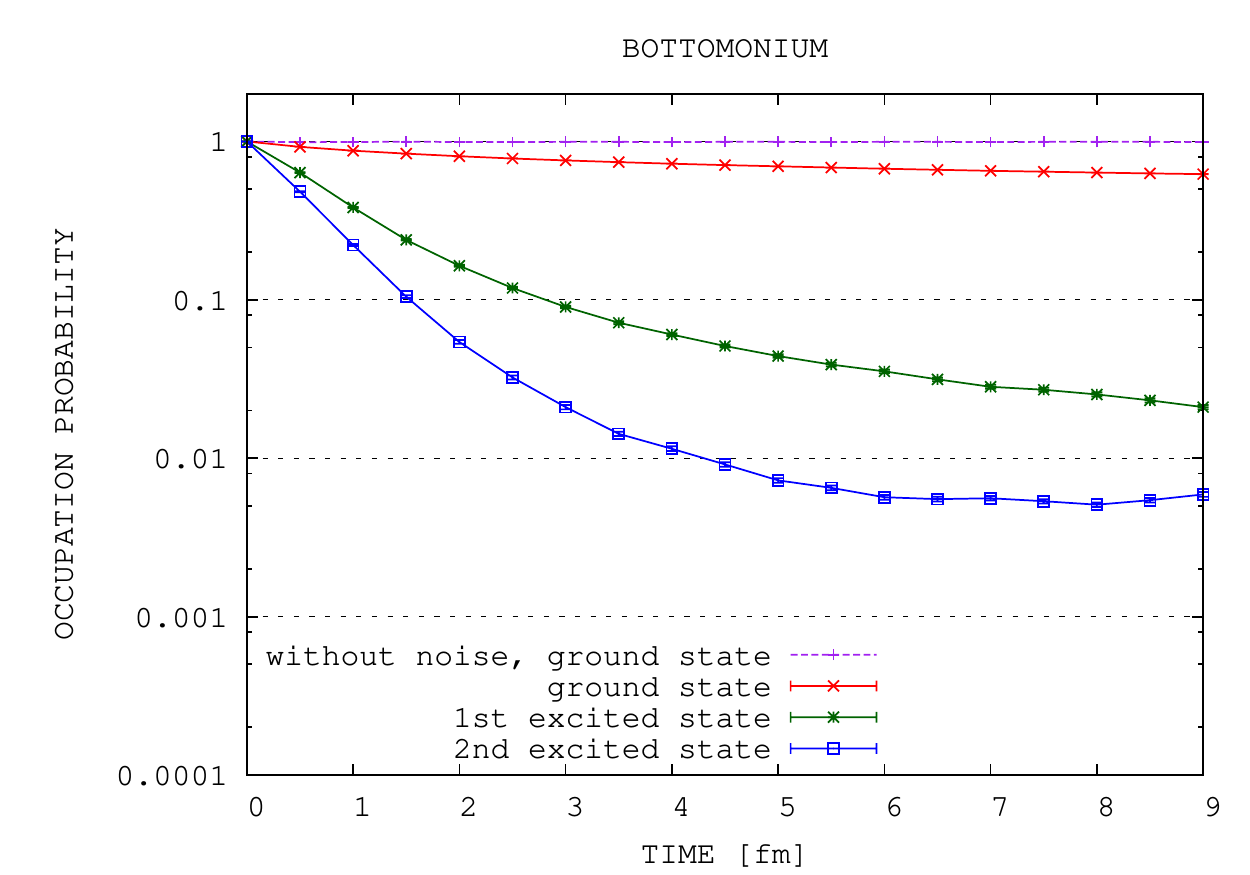}\\
\includegraphics[clip,width=9.0cm,bb= 0 0 380 270]{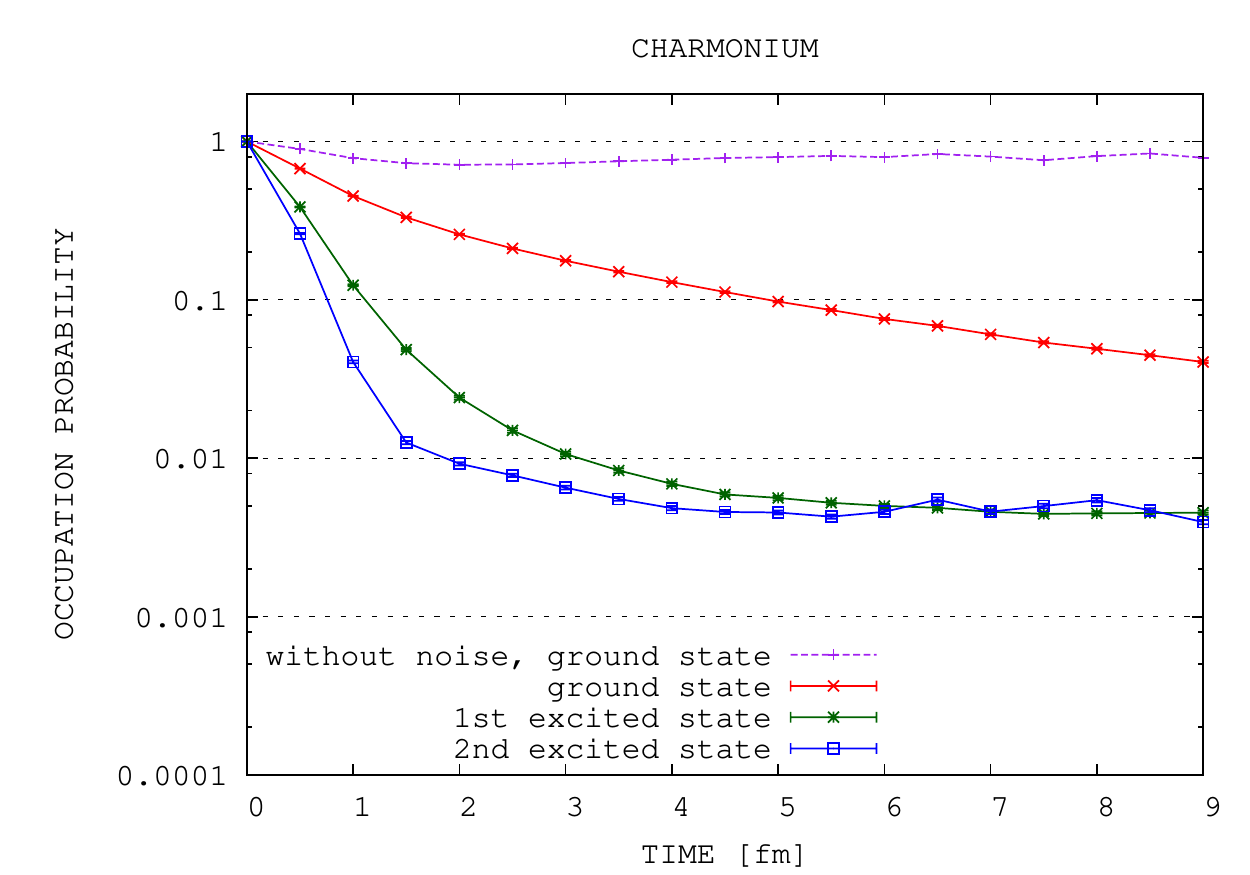}
\end{tabular}
\caption{
Time evolution of the occupation probability of quarkonium bound states (the ground, the first excited, and the second excited state) in the stochastic potential model in a Bjorken-expanding QGP.
Both the initial states and the projected states are the bound states in the vacuum Cornell potential.
The upper figure shows the calculation for bottomonium and the lower figure shows that for charmonium.
For comparison purposes, we also plot the probability of the ground state from an evolution only with the Debye screened potential, i.e. without noise (dashed lines).
}
\label{fig:su3expand}
\end{center}
\end{figure}

Figure \ref{fig:su3expand} shows the occupation probability of the ground, first excited, and second excited states of bottomonium and charmonium starting from each state.
The excited states are more extended than the ground states so that their occupation probabilities decrease faster than that of the ground states.
It is also confirmed that the occupation numbers of charmonia decrease faster than those of bottomonia because the charmonia are generally larger than the bottomonia.
In this figure, we also compare the ground state occupations by the stochastic simulation and those by the Schr\"odinger equation without noise (and thus without the imaginary part).
The former simulates both the static and dynamical in-medium effects in the potential while the latter sees how the static effect, namely the Debye screening, affects the quarkonium occupations.
It clearly demonstrates that the dynamical in-medium effect is much more important in the quarkonium dissociation.

\begin{figure}
\begin{center}
\includegraphics[clip,width=8.5cm,bb= 0 0 450 380]{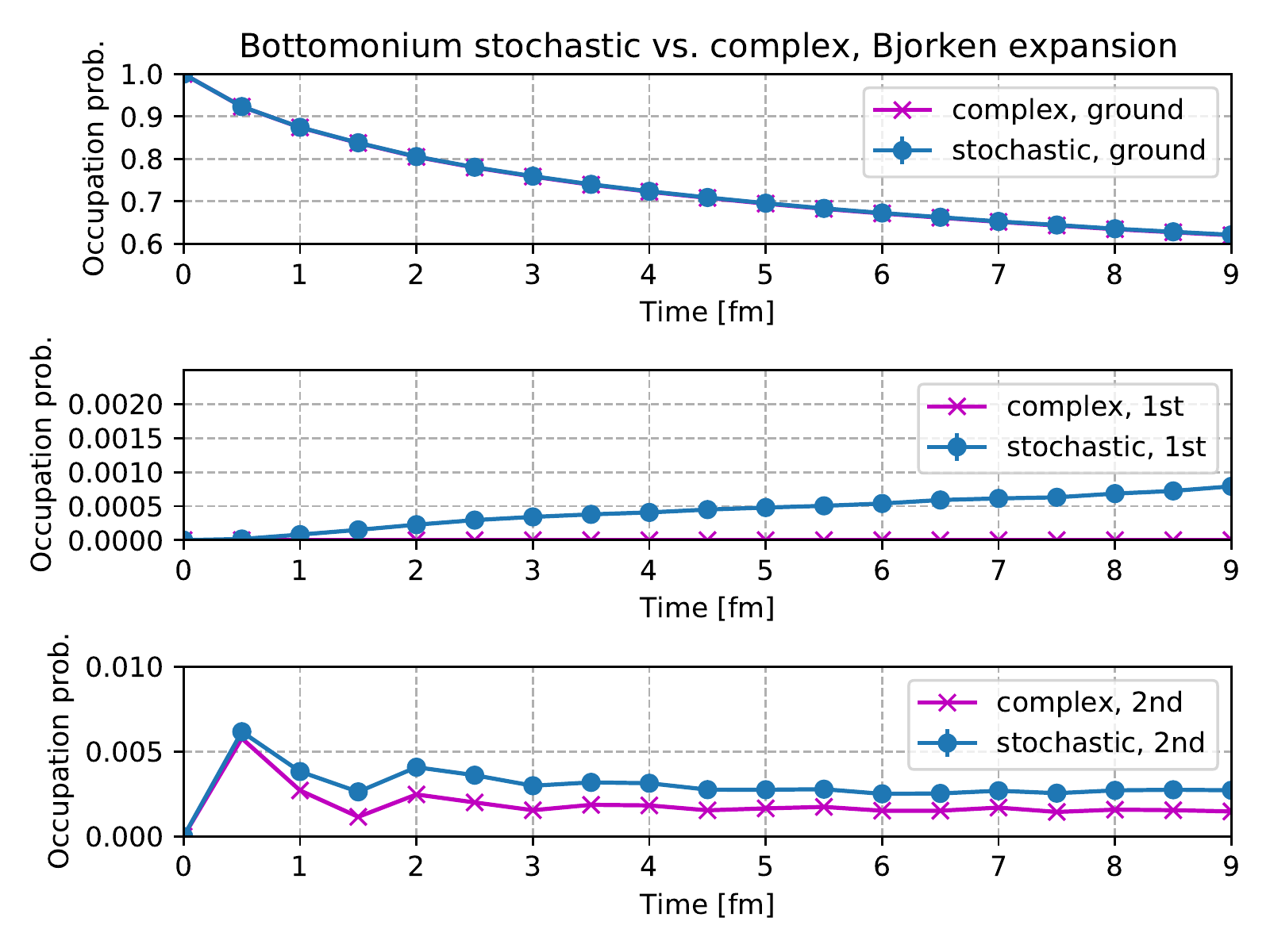}
\caption{
Time evolution of occupation probability of the bottomonium bound states (the ground, the 1st excited, and the 2nd excited states) in a Bjorken expanding QGP.
The initial condition is the ground state.
The bound states are calculated in the vacuum Cornell potential.
}
\label{fig:su3sc}
\end{center}
\end{figure}

In Fig.~\ref{fig:su3sc}, we compare the simulations by the stochastic and complex potentials for bottomonia in the Bjorken expanding QGP.
As in the previous section, we confirm that the survival probabilities of bottomonium ground states are essentially the same within the errorbar.
However, the transition probabilities from the singlet ground state to the singlet excited states show some differences.
In particular, the complex potential cannot induce transitions between different parity states (from the ground to the first excited state), which are possible in the stochastic potential model with $N_c\neq 2$.
Furthermore, the transition between the same parity states (from the ground to the second excited states) is underestimated about 50\% by the complex potential.
If the future modeling of the quarkonium in the heavy-ion collisions requires this level of precision, the stochastic potential model must be adopted.

\subsection{Spatial distribution and density matrix}\label{sec:simulation_nandrho}
Finally, let us show how the spatial distribution and density matrix develops in the stochastic Schr\"odinger equation \eqref{eq:sse3}.
As in Sec.~\ref{sec:simulation_bjorken}, we perform the simulation for the bottomonium in the Bjorken expanding QGP starting from the bottomonium ground state.
Here, we show our results on spatial distribution and density matrix for the relative coordinate $r$ of the bottomonium.

\begin{figure}
\begin{center}%
\includegraphics[clip, bb= 0 0 360 250, width = 0.4\textwidth]{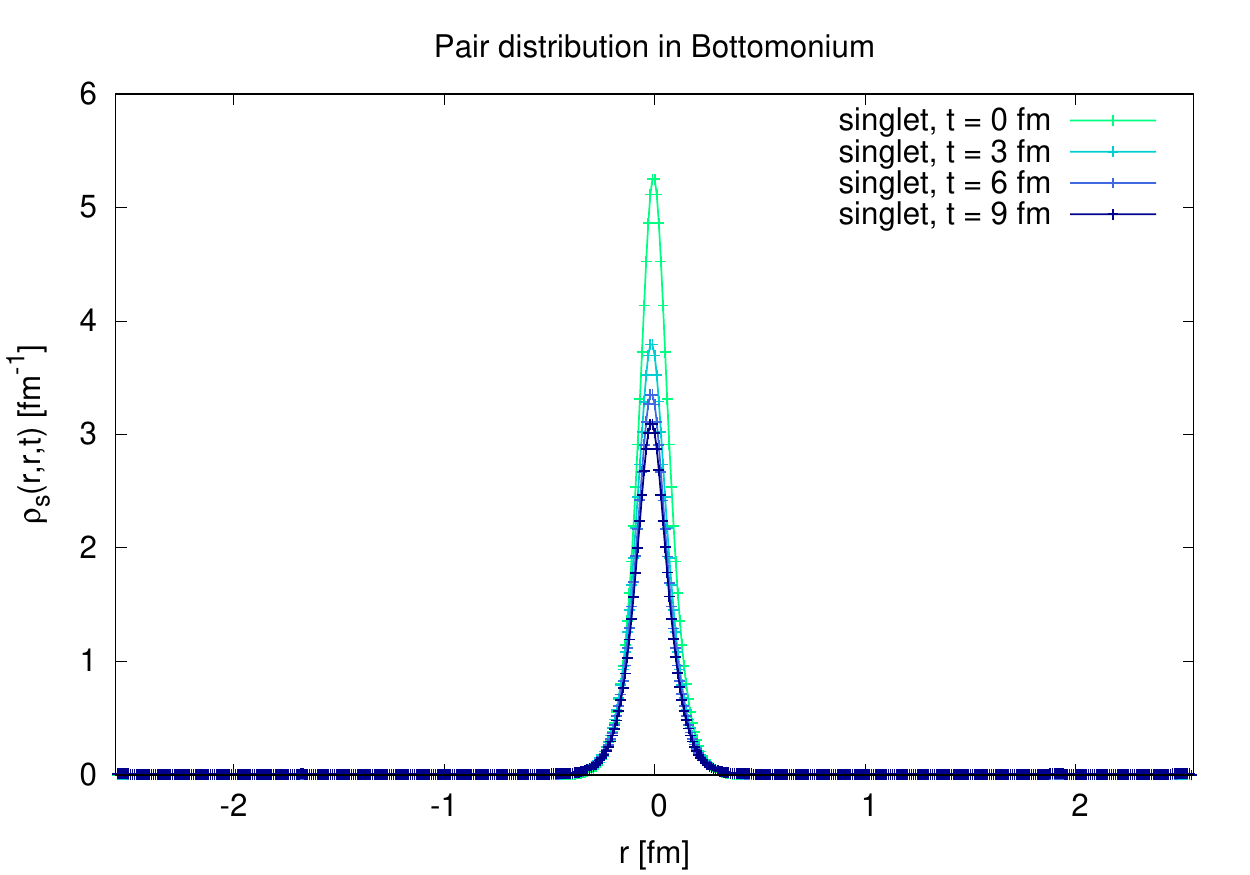}\\
\includegraphics[clip, bb= 0 0 360 250, width = 0.42\textwidth]{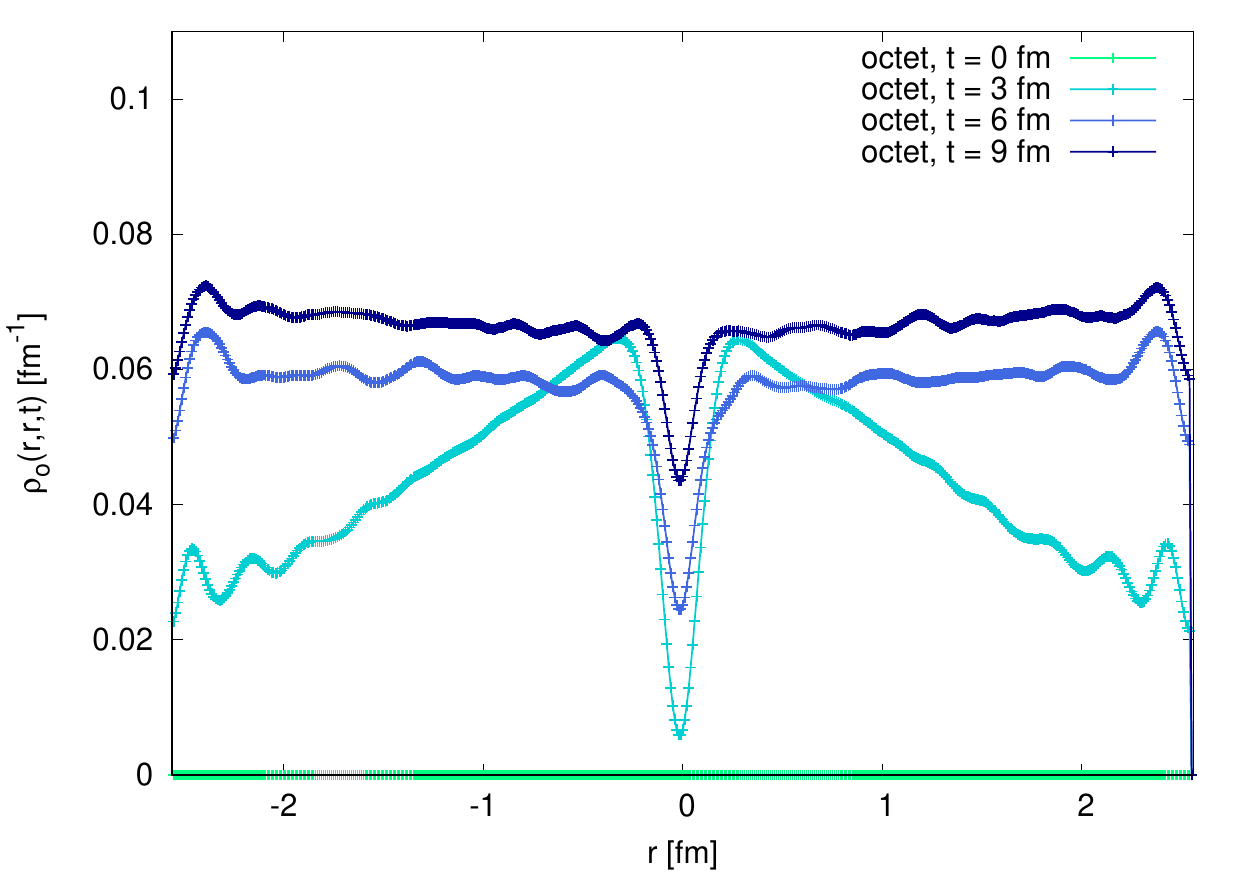}
\caption{
Spatial distribution of the bottom quark pair in the color singlet (upper) and octet (lower) states in a Bjorken expanding QGP at different times $t=0, 3, 6, 9$ fm.
The initial condition is the ground state in the vacuum Cornell potential.
}
\label{fig:distribution}
\end{center}
\end{figure}

Figure \ref{fig:distribution} demonstrates the spatial distribution of the bottom quark pair in the color singlet and octet states at different times $t=0, 3, 6, 9$ fm.
From this figure, we observe that the color singlet distribution does not get spatially extended while the color octet distribution does.
This is consistent with our previous findings that the complex potential simulation works reasonably well because the de-excitation from the color octet to the singlet is suppressed when $N_c$ is large.
We also find that the distance between the bottom quark pair in the color octet states well exceeds the medium length scales, which we take $1/m_D = l_{\rm corr} = 1/T$, and even reaches the spatial boundary of the numerical simulation.
It casts some doubt on the applicability of the dipole approximation for the interaction between quarkonium and in-medium gluons.
In principle, we can check whether the occupation probability calculated by the stochastic potential model in the dipole limit stays within acceptable deviation from the original one.
In the dipole limit, the noise terms are approximated by (see Appendix \ref{app:lindblad} for derivation)
\begin{subequations}
\label{eq:hamiltonian_dipole}
\begin{align}
&\Theta(\bm{r},t) = \frac{\bm r}{2}\cdot \bm f^a(t) \left(t^a\otimes 1 + 1\otimes t^{a*}\right),\\
&\langle f^a_i(t)f^b_j(t')\rangle =  -\frac{\bm\nabla^2 D(0)}{3}\delta(t-t')\delta^{ab}\delta_{ij}.
\end{align}
\end{subequations}

In the simplified U(1) case, it is numerically indicated \cite{Sharma:2019xum} that the small-$r$ expansion is reliable for the calculation of survival probability as long as the bound state is smaller than the noise correlation length $\sim l_{\rm corr}$.
In such a case, it is expected that the excitation rate of a bound state is reliably approximated by the small-$r$ expansion.
Even if the evolution of the disturbed wave function evolves incorrectly at large distances, the calculation of the survival probability will be more or less unaffected because the chance of de-excitating back to the original bound state is rather low, as suggested by the comparison to the complex potential simulations.
This is how we interpret the effectiveness of small-$r$ expansion for the U(1) case.
Therefore, we also expect that the small-$r$ expansion for the SU(3) case is more reliable than the U(1) case because the probability of returning back to the color singlet bound state is even smaller due to the color structure.

\begin{figure*}
\begin{center}
\includegraphics[clip, bb= 50 0 330 250, width=0.32\textwidth]{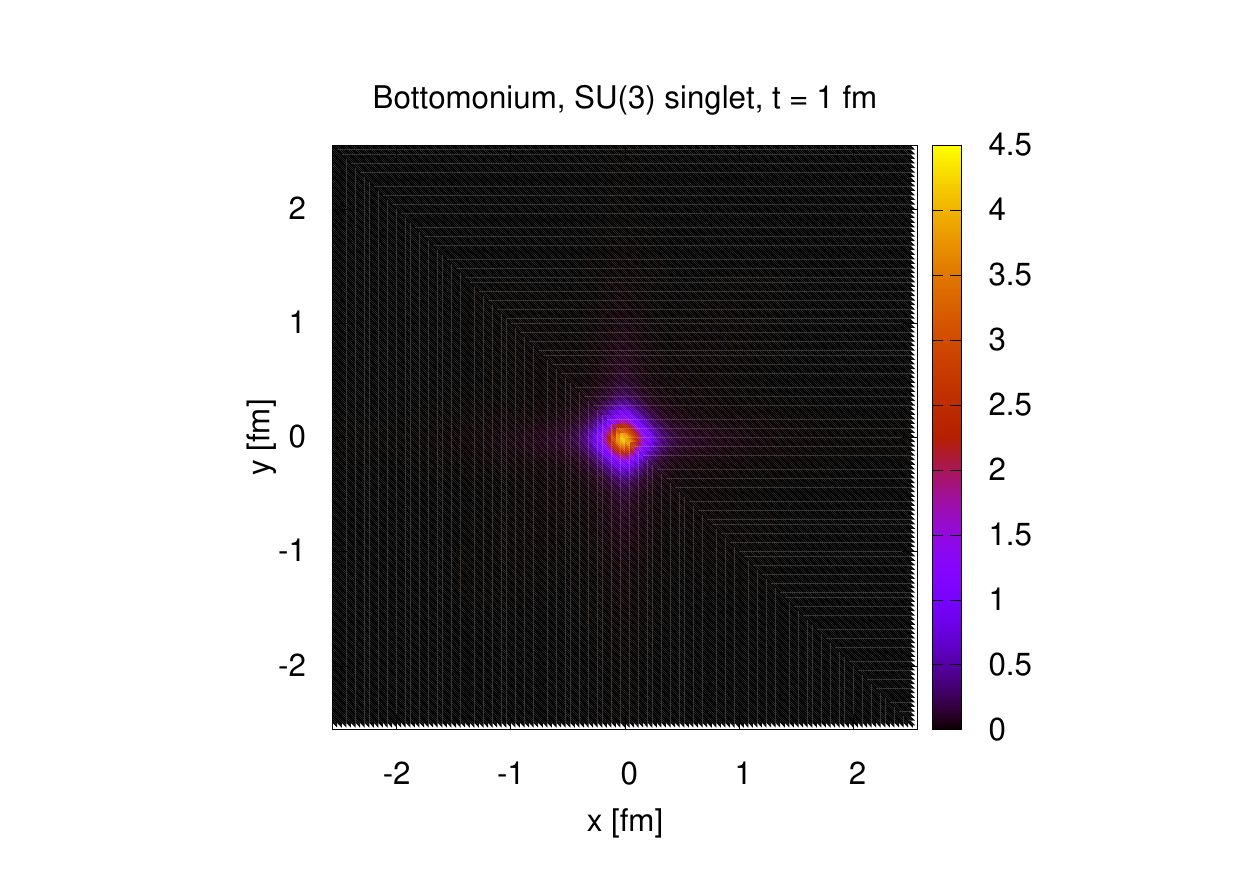}
\includegraphics[clip, bb= 50 0 330 250, width=0.32\textwidth]{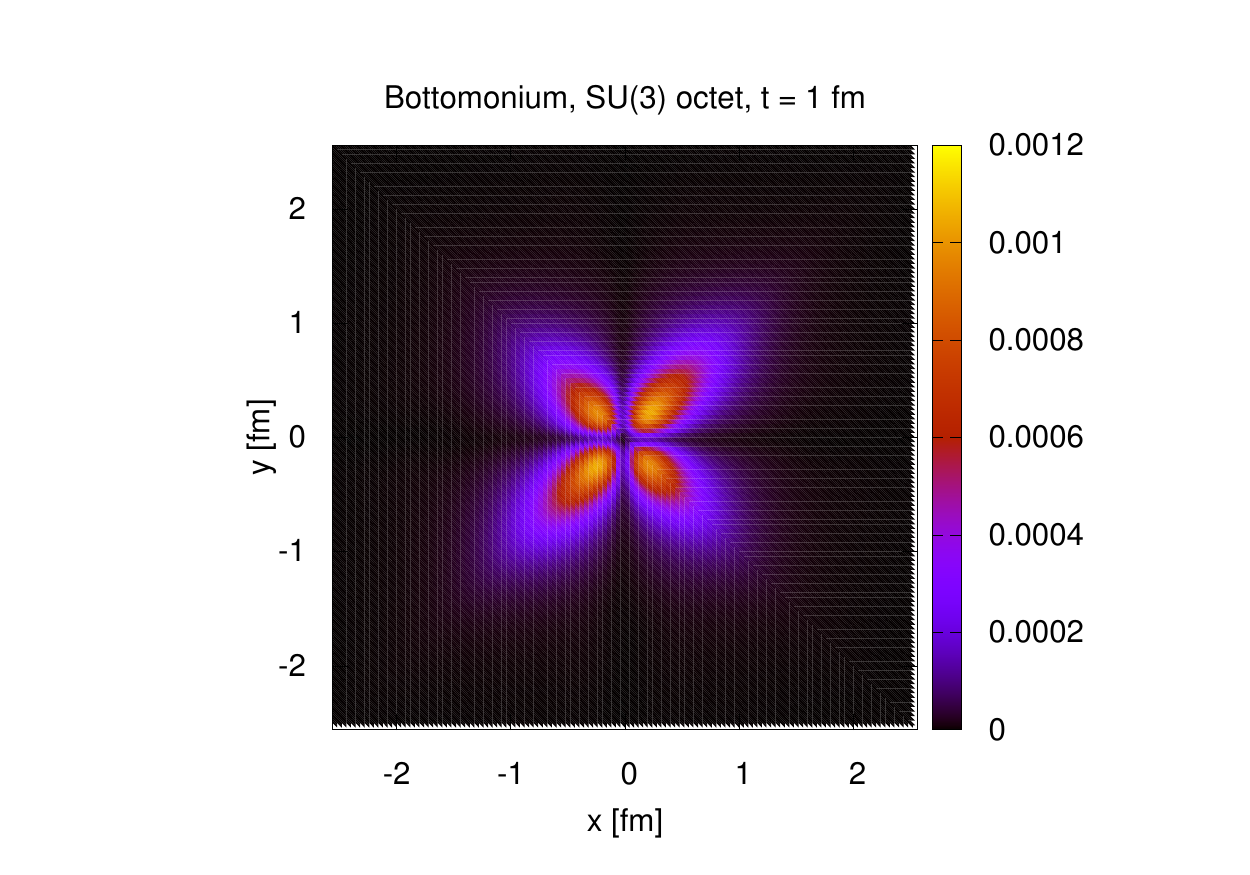}
\includegraphics[clip, bb= 50 0 330 250, width=0.32\textwidth]{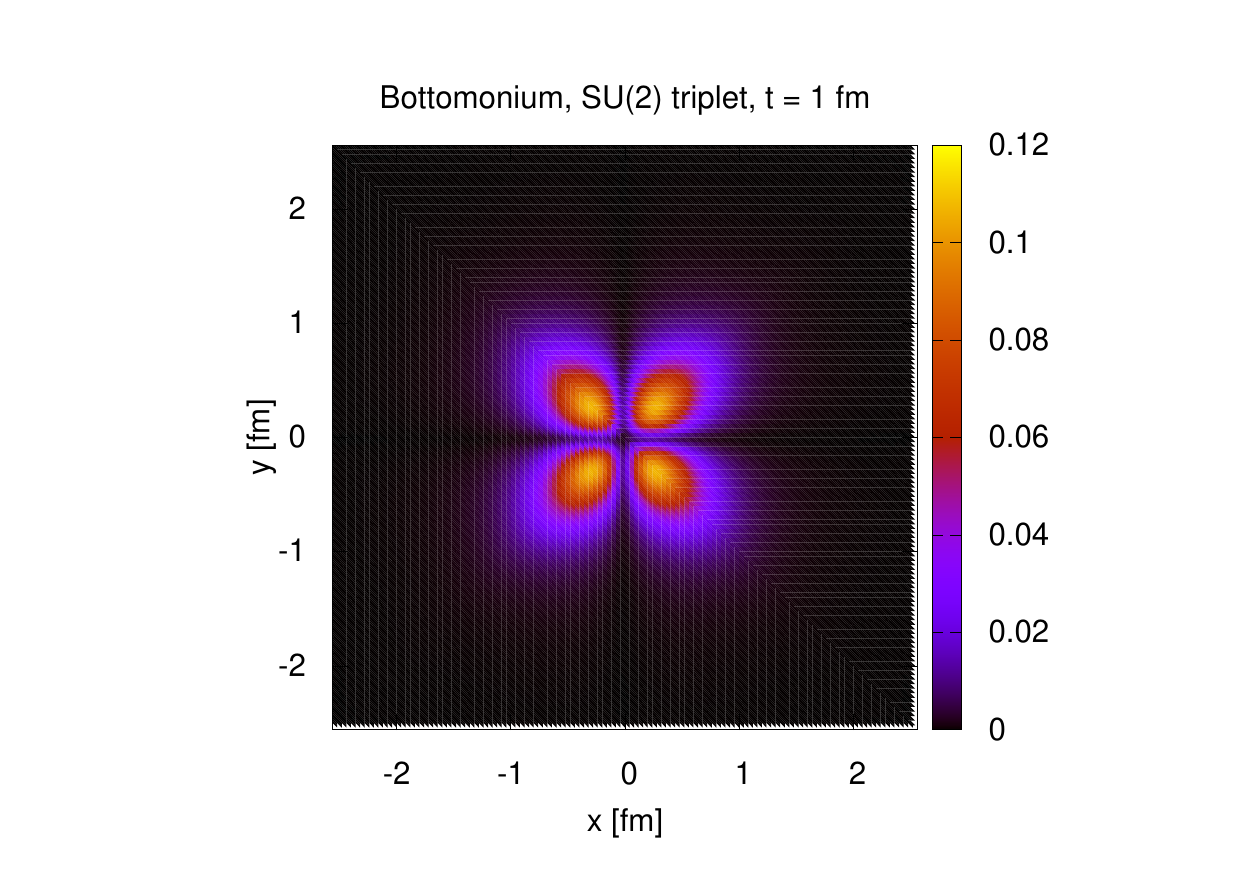}\\
\includegraphics[clip, bb= 50 0 330 250, width=0.32\textwidth]{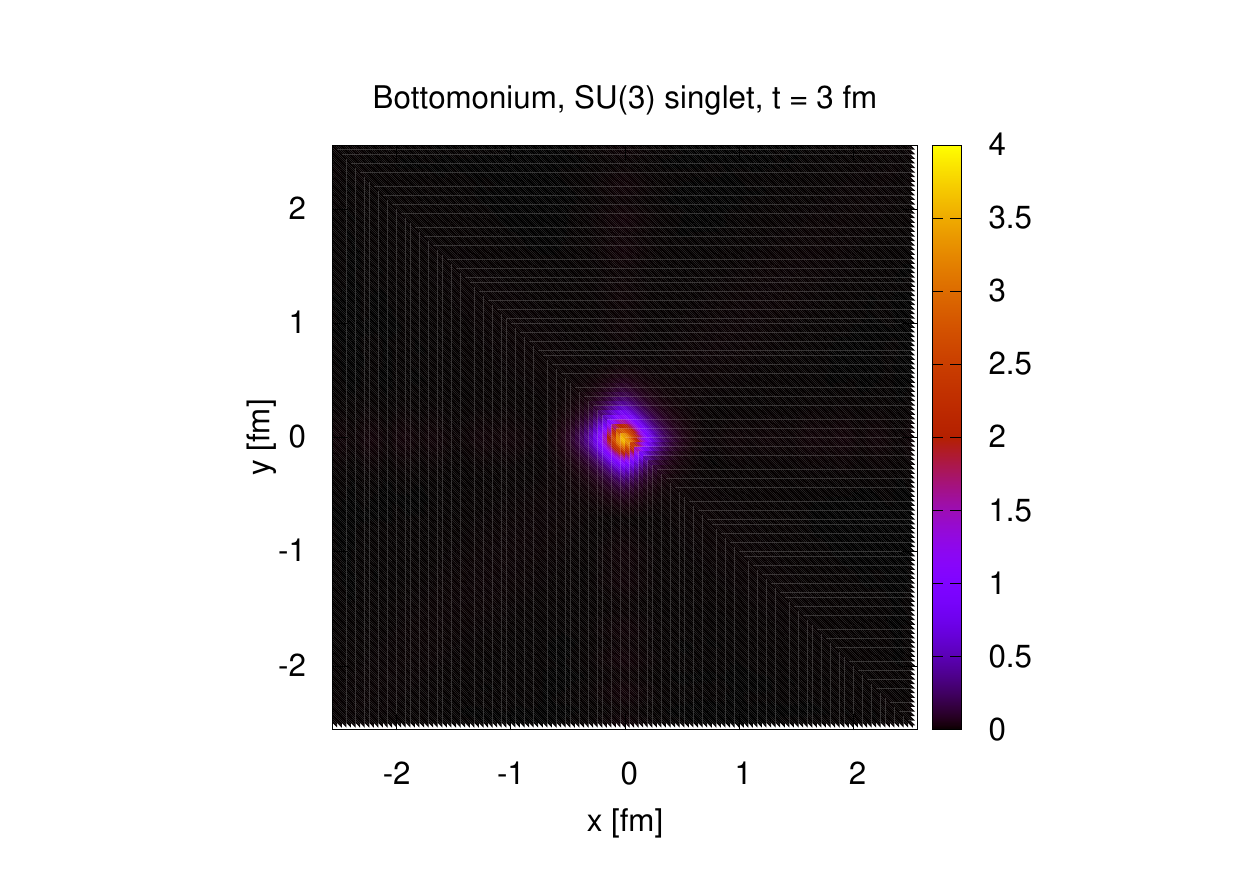}
\includegraphics[clip, bb= 50 0 330 250, width=0.32\textwidth]{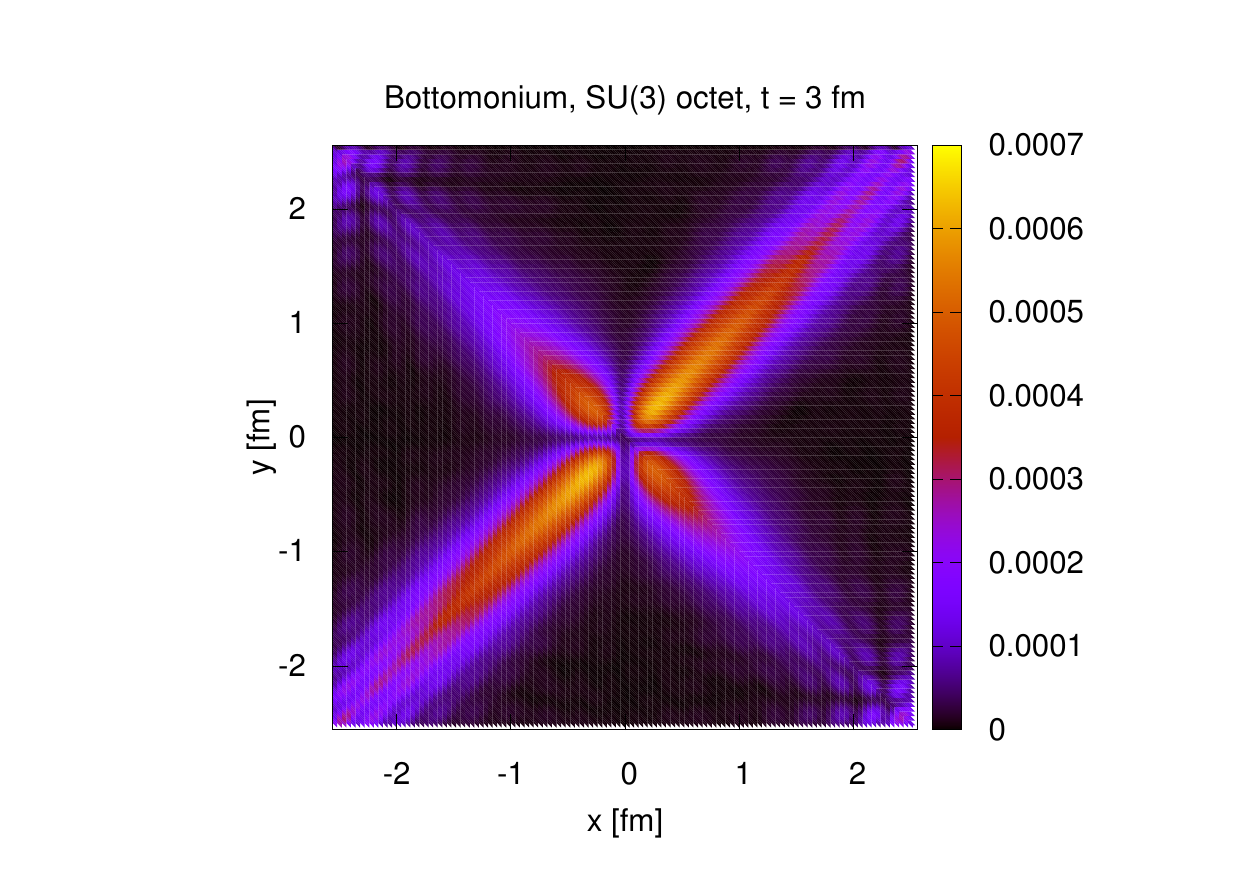}
\includegraphics[clip, bb= 50 0 330 250, width=0.32\textwidth]{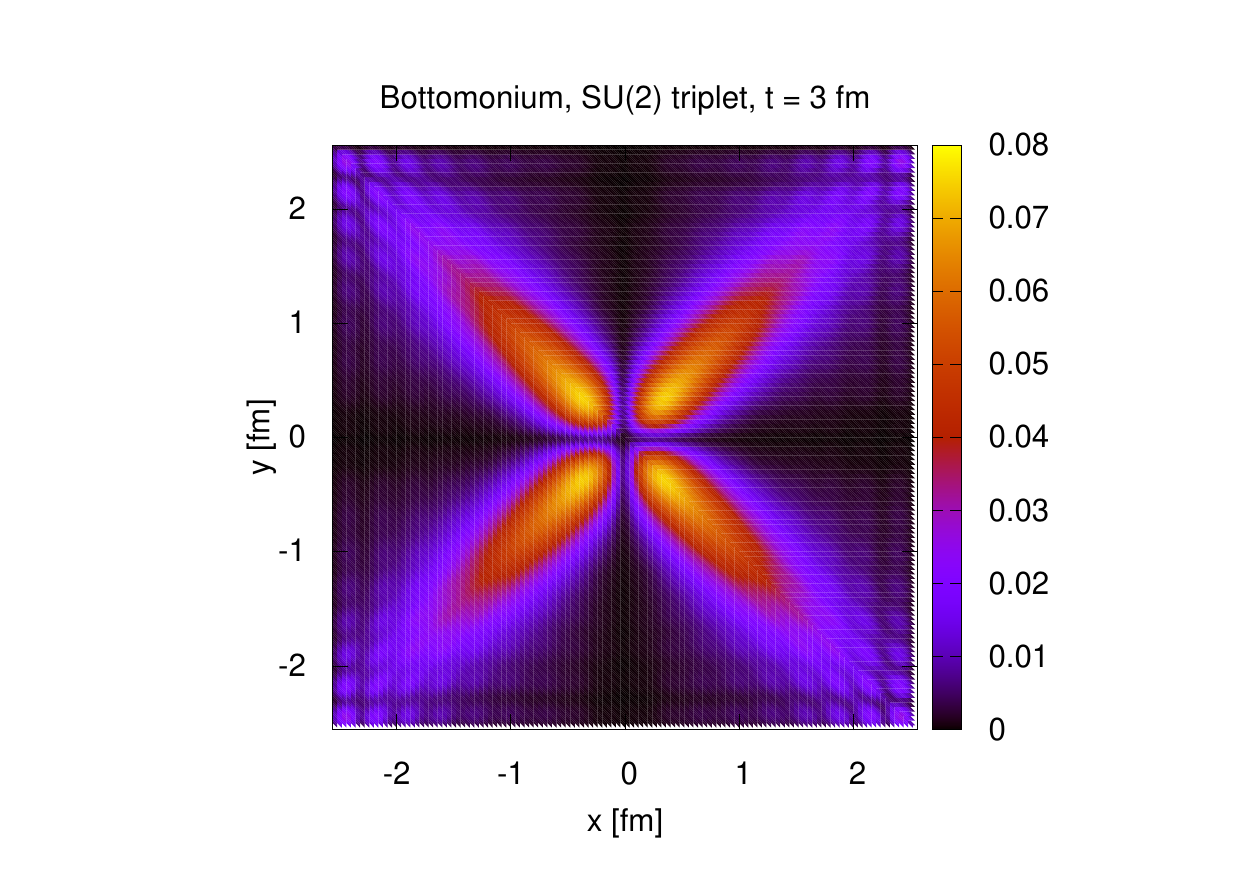}\\
\includegraphics[clip, bb= 50 0 330 250, width=0.32\textwidth]{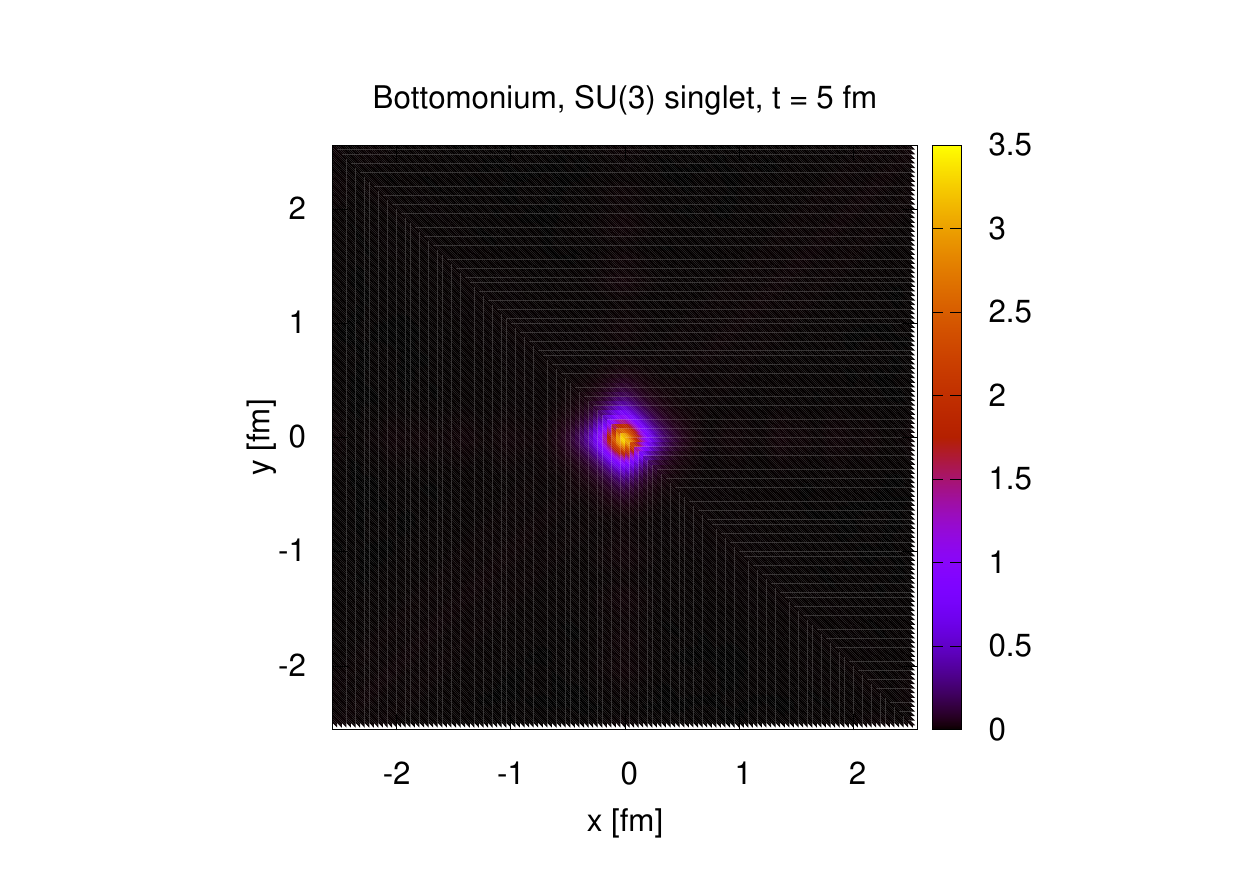}
\includegraphics[clip, bb= 50 0 330 250, width=0.32\textwidth]{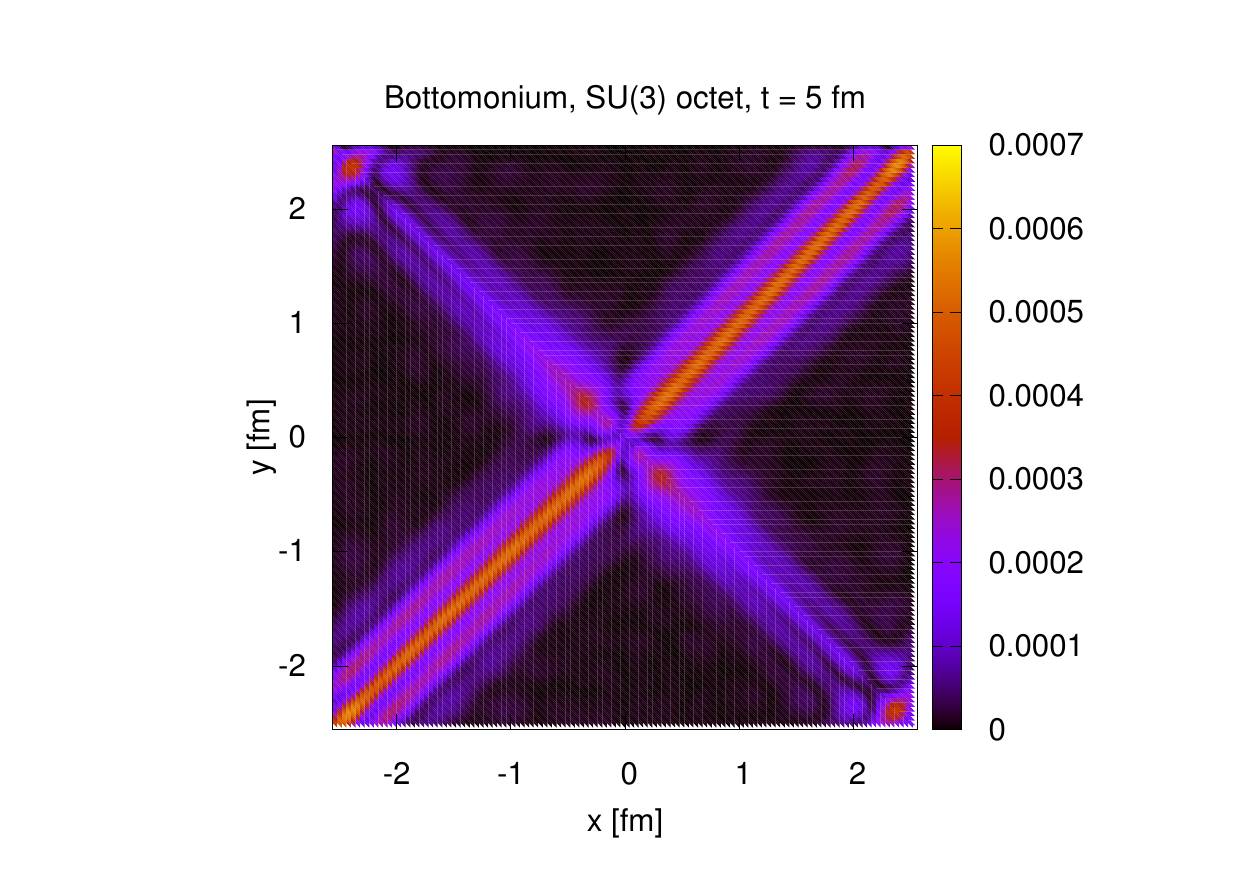}
\includegraphics[clip, bb= 50 0 330 250, width=0.32\textwidth]{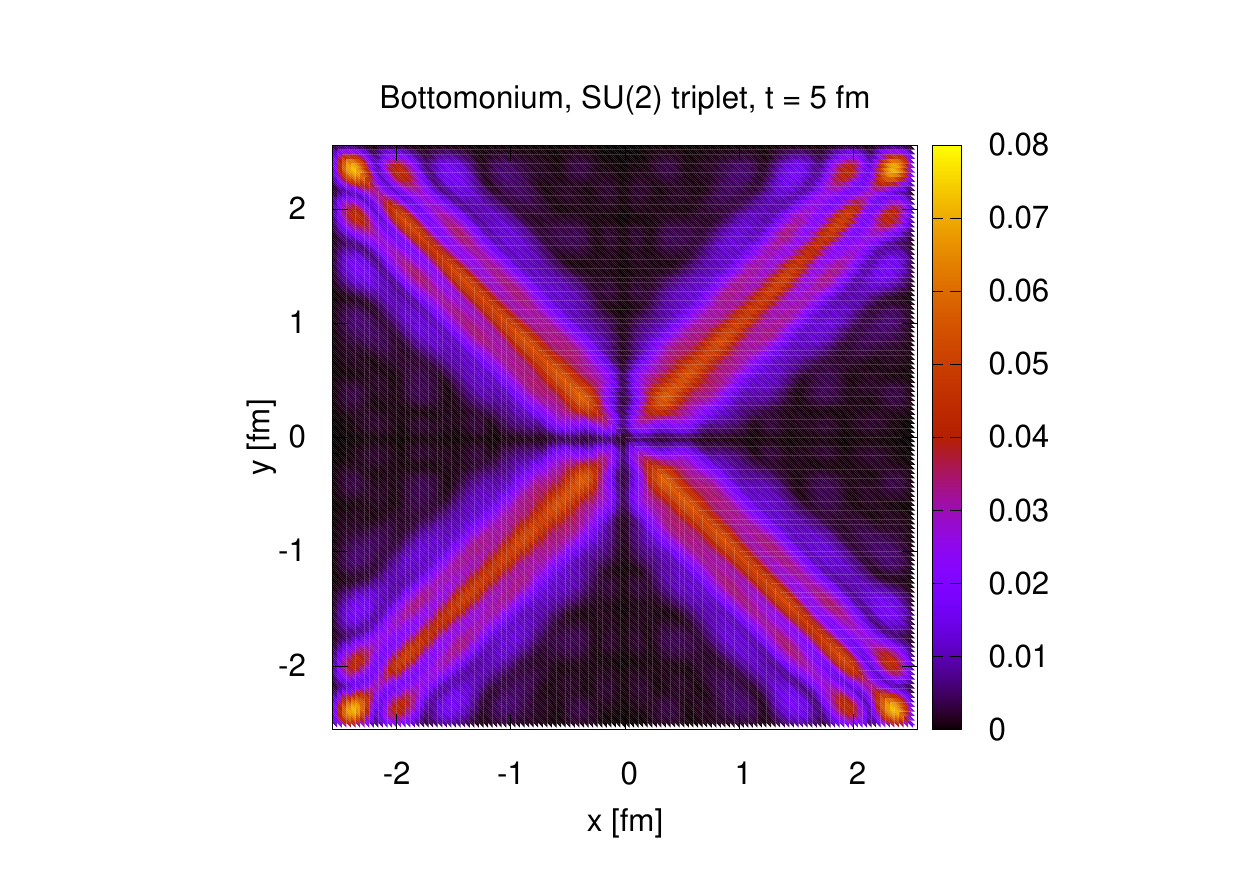}\\
\includegraphics[clip, bb= 50 0 330 250, width=0.32\textwidth]{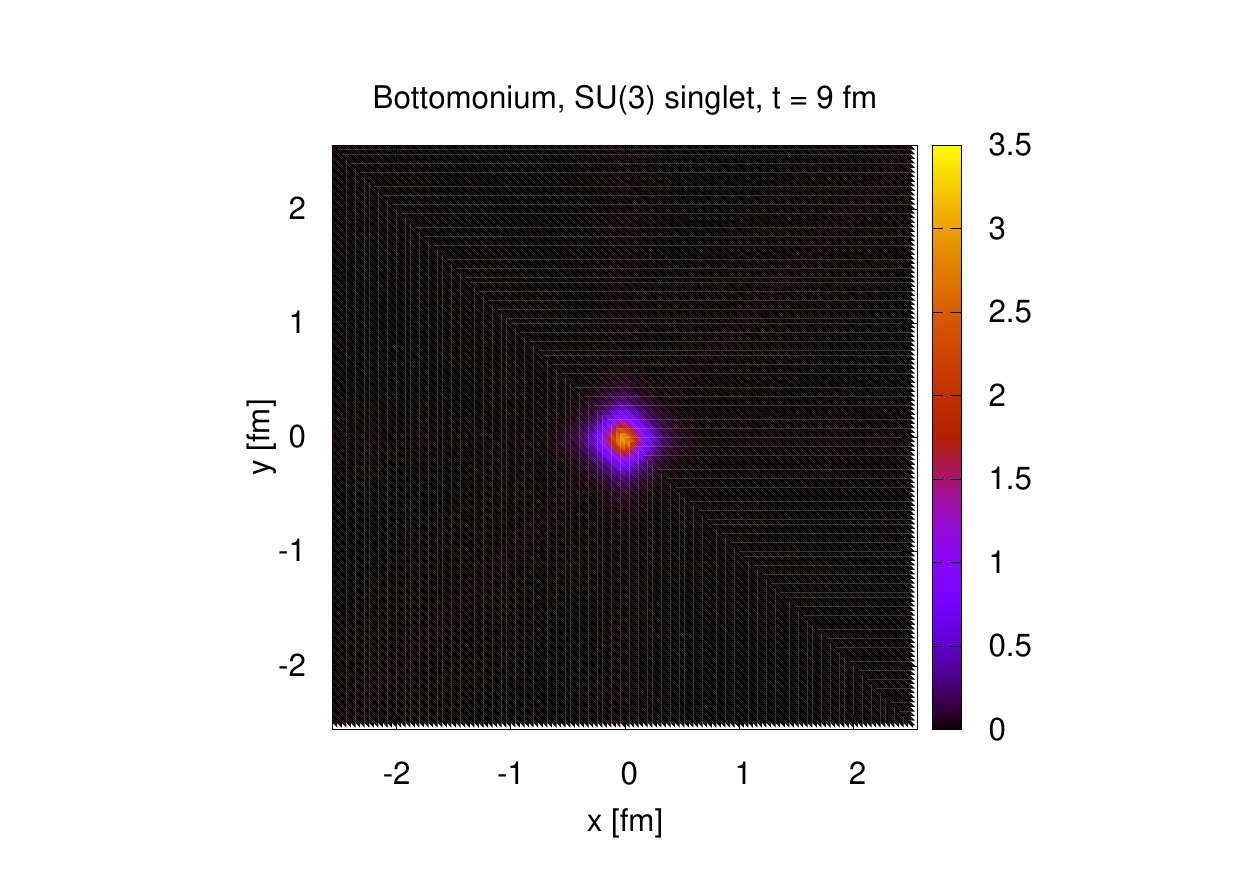}
\includegraphics[clip, bb= 50 0 330 250, width=0.32\textwidth]{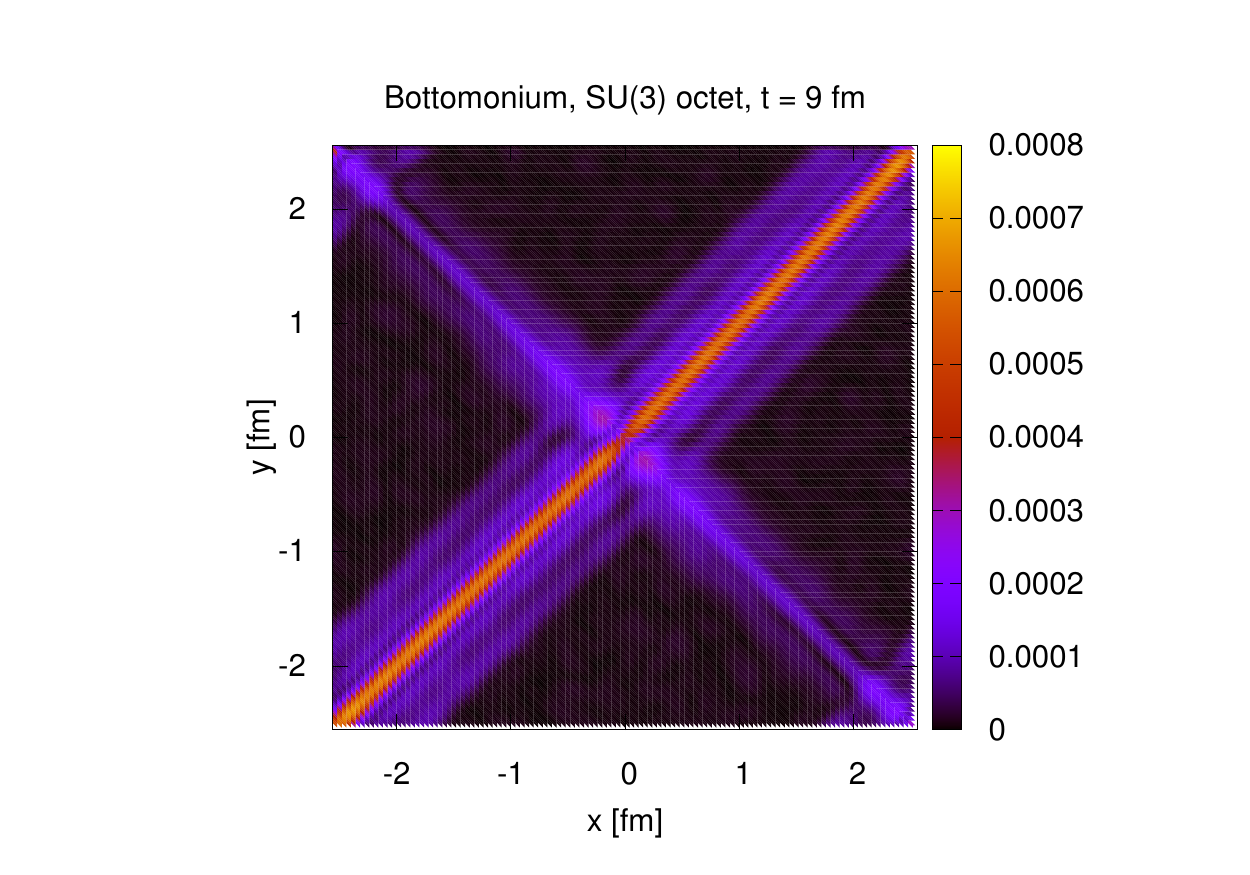}
\includegraphics[clip, bb= 50 0 330 250, width=0.32\textwidth]{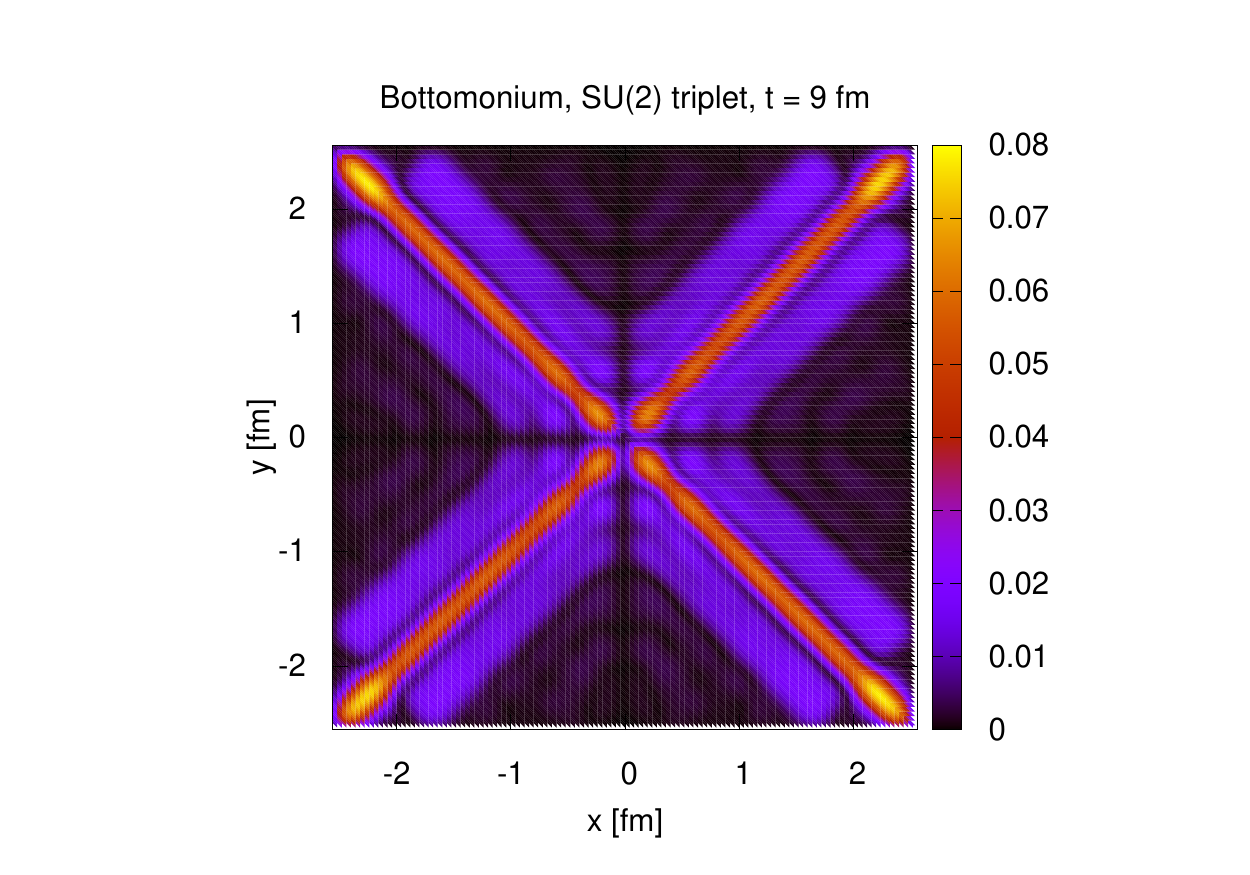}\\
\caption{
Density matrices of bottomonium calculated by the SU(3) and SU(2) stochastic potential models in a Bjorken expanding QGP.
The initial state is the ground state in the vacuum Cornell potential.
Plotted are the absolute values of density matrices $|\rho(x,y,t)| \ {\rm fm}^{-1}$ at different times $t=1,3,5,9$ fm, for the color SU(3) singlet (left) and octet (center) states, and for the color SU(2) triplet states (right).
}
\label{fig:initparity}
\end{center}
\end{figure*}

In Fig.~\ref{fig:initparity}, we show the absolute value of bottomonium density matrix in the octet/triplet sector for the color SU(3)/SU(2) stochastic potential model.
We plot the results at different times $t=1,3,5,9$ fm in our simulation on the Bjorken expansion.
The result of SU(3) shows the diagonalization process of density matrix due to the decoherence.
In the color octet sector, first population is supplied from the ground state in a color dipole-like configuration.
Such excitation from the color singlet continues while the octet distribution extends because of the repulsive force.
At the same time, the coherence of octet states is gradually lost due to the decoherence.
Although the off-diagonal components at $x+y\simeq 0$ remain finite, classical particle description is more or less applicable at $t \sim 5$ fm in the color-octet channel.
In the color singlet, the attractive potential localizes the wave functions around the origin, where the classical particle description is hindered.

In contrast, the color triplet density matrix of SU(2) case does not diagonalize but a structure ``X" shows up even at late times.
This structure is protected by the event-by-event symmetry of the spinor conjugation as explained in Sec.~\ref{sec:symmetry_S}.
Since the initial condition is the singlet ground state, which is parity even, $S=-1$ is conserved and the triplet state must be parity odd.
The odd parity relates the diagonal  and off-diagonal parts $\rho_t(\bm r, \bm r) = \rho_t(-\bm r, -\bm r)= -\rho_t(\bm r, -\bm r)= -\rho_t(-\bm r, \bm r) $ of the density matrix in the triplet.
In the color-triplet channel, the conventional picture of classic particle must be extended.
Even in the classical picture, the pair of particles with relative position $\bm r$ is a superposition of heavy quark (antiquark) being at $\bm R + \bm r/2$ ($\bm R-\bm r/2$) and heavy quark (antiquark) being at $\bm R - \bm r/2$ ($\bm R+\bm r/2$).
It is only when we measure the local heavy quark number that this superposition states shrinks to one of the above two possibilities. 
The color singlet density matrix of SU(2) case is similar to the SU(3) case so that we do not show it here.

\section{Conclusion}\label{sec:conclusion}
In this paper, we analyze the symmetries of the SU($N_c$) stochastic potential model in detail.
The symmetries discussed in this paper are global SU($N_c$), parity, and charge conjugation.
All these symmetries are symmetries of the total system (QCD including the heavy quarks) and are expected in the open system description of quarkonia in the QGP environments if the environment possesses these symmetries.
These symmetries are realized in the quarkonium system after taking the event averages.

In the case of $N_c=2$, we find an ``event-by-event" symmetry and name it ``spinor conjugation" $S$.
Spinor conjugation $S$ for quarkonium is defined in Eq.~\eqref{eq:spinorconj_tensor} or \eqref{eq:spinorconj_matrix}.
At the fundamental level, it is a symmetry of the SU(2) gauge theory and exchanges heavy quark and antiquark with color and spin assignments specified in \eqref{eq:spinorconj_dirac}. 
Since the gauge fields are invariant under $S$, it is realized as an event-by-event symmetry in the SU(2) stochastic potential model.
An interesting physical consequence of this symmetry is that the density matrix for the relative coordinates has a structure ``X" and the off-diagonal components do not decay.
This is in sharp contrast to the usual scenario of decoherence, where the density matrix diagonalizes.
Physically, the structure ``X" represents a superposition of two classical states: heavy quark (antiquark) being at $\bm R + \bm r/2$ ($\bm R-\bm r/2$) and heavy quark (antiquark) being at $\bm R - \bm r/2$ ($\bm R+\bm r/2$).
These states are, as it were, ``decoherence-free pair"\footnote{
Although these states do not fit into the definition of the ``decoherence-free subspace" \cite{Schlosshauer:2019ewh}, the physical consequence is similar.
In our case, if $|\psi(t_0)\rangle=S|\phi(t_0)\rangle$ holds initially, it holds at later time as well, i.e. $|\psi(t)\rangle=S|\phi(t)\rangle$, so that $|\psi\rangle$ and $|\phi\rangle$ stay in pair even in the presence of decoherence.
}.

We also show the numerical simulations of the SU($N_c$) stochastic potential model in a static and Bjorken expanding QGP.
Some of our findings are listed as follows:
\begin{itemize}
\item Dynamical effect (decoherence) is more important than the static effect (screening) [Fig.~\ref{fig:su3expand}].
\item Complex potential is precise enough for the calculation of survival probability while it is not for the transition probability [Figs.~\ref{fig:ft3com} and \ref{fig:su3sc}].
\item {Spatial distribution extends rapidly (only) in the color octet states [Fig.~\ref{fig:distribution}].}
\item {Diagonalization for the color octet takes place about $t\sim 5$ fm for color SU(3), which however does not happen for color SU(2) due to the constraint by the spinor conjugation [Fig.~\ref{fig:initparity}].}
\end{itemize}
If the dissipative effects can be ignored (see \cite{Akamatsu:2018xim, Miura:2019ssi} for recent studies in this direction) and if the observed suppression of $\Upsilon(n{\rm S})$ is determined dominantly by its survival probability, our simulation indicates that the Schr\"odinger equation with the complex potential \cite{Strickland:2011mw, Strickland:2011aa, Krouppa:2015yoa, Krouppa:2016jcl, Krouppa:2017jlg, Islam:2020gdv} is a reasonably useful effective description for at least phenomenological purposes.

\acknowledgments
We thank Alexander Rothkopf for the discussions in the early stage of this work.
The work of Y.A. is supported by JSPS KAKENHI Grant Number JP18K13538.
M.A. is supported in part by JSPS KAKENHI Grant Number JP18K03646.

\appendix
\section{Lindblad equations for the SU($N_c$) stochastic potential}\label{app:lindblad}
The evolution of the reduced density matrix $\rho(t)$ must conserve its trace ${\rm Tr}\rho(t) = 1$ and operator positivity $\rho(t)\geq 0$.
In the Markov limit, such evolution equation must take the following form \cite{Gorini:1975nb, Lindblad:1975ef}, known as the Lindblad form
\begin{align}
\frac{d}{dt}\rho(t) = -i[H, \rho] + \sum_i \left ( L_i \rho L_i^{\dagger} - \frac{1}{2}L_i^{\dagger}L_i\rho - \frac{1}{2}\rho L_i^{\dagger}L_i \right ),
\end{align}
where the $L_i$s are the Lindblad operators and the term $\sum_i \left ( L_i\rho L_i^{\dagger} - \frac{1}{2}L_i^{\dagger}L_i\rho - \frac{1}{2}\rho L_i^{\dagger}L_i \right )$ is called the dissipator.

\subsection{In the tensor basis}\label{app:lindblad_tensor}

The SU($N_c$) stochastic potential model \eqref{eq:hamiltonian} and \eqref{eq:sunnoiseproperty} in the Sec.~\ref{sec:intro} is one of the stochastic unravelling methods for the Lindblad master equation \cite{Akamatsu:2014qsa}
\begin{align}
\label{eq:Lindblad}
&\frac{\partial}{\partial t}\rho(t)
= -i\left[\frac{\bm p^2}{M} + V(\bm r)(t^a\otimes t^{a*}), \rho\right] \\
&\quad + \int\frac{d^3 k}{(2\pi)^3}\tilde D(\bm k) \left(V^a_{\bm k}\rho V_{\bm k}^{a\dagger}
- \frac{1}{2}V_{\bm k}^{a\dagger}V^a_{\bm k}\rho - \frac{1}{2}\rho V_{\bm k}^{a\dagger}V^a_{\bm k}\right), \nonumber
\end{align}
where $\tilde D(\bm k) = \int d^3 x D(\bm x) e^{-i\bm k\cdot\bm x}>0$ and $V^a_{\bm k}$ describes the rotation of heavy quark color in a scattering with momentum transfer $\bm k$
\begin{align}
V^a_{\bm k} = e^{i\bm k\cdot\bm r/2}(t^a\otimes 1) - e^{-i\bm k\cdot\bm r/2}(1\otimes t^{a*}).
\end{align}
Here, unlike in the main text, $\bm r$ represents an operator for the relative coordinate while $\bm k$, $\bm x$, and $\bm y$ are vector-valued parameters.
It is easy to see the connection to the stochastic potential by expressing the dissipator of the Lindblad equation in the configuration space
\begin{align}
\mathcal D(\rho) &= \int d^3x d^3y D(\bm x- \bm y) \nonumber \\
& \quad\times\left(
V^a_{\bm y}\rho V^{a\dagger}_{\bm x} - \frac{1}{2}V^{a\dagger}_{\bm x}V^a_{\bm y}\rho
-\rho \frac{1}{2}V^{a\dagger}_{\bm x}V^a_{\bm y}
\right),
\end{align}
with now $V^a_{\bm x}$ rotates the heavy quark color at $\bm x$
\begin{align}
V^a_{\bm x} = \delta\left(\bm x- \frac{\bm r}{2}\right) (t^a\otimes 1) - \delta\left(\bm x + \frac{\bm r}{2}\right)(1\otimes t^{a*}).
\end{align}
An important property is that the Lindblad operator is hermite $V^a_{\bm x} = V^{a\dagger}_{\bm x}$, which allows us to rewrite the dissipator in the following stochastic representation
\begin{subequations}
\begin{align}
&H_{\Theta} = \frac{\bm p^2}{M} + V(\bm r)(t^a\otimes t^{a*}) + \int d^3 x \theta^a(\bm x, t)V^a(\bm x), \\
&\langle\theta^a(\bm x,t)\theta^b(\bm x', t')\rangle = D(\bm x-\bm x')\delta(t-t')\delta^{ab}.
\end{align}
\end{subequations}

\subsection{In the singlet-octet basis}\label{app:lindblad_so}
When one is interested only in the singlet and octet occupation\footnote{
Here we do not use the assumption made in Sec.~\ref{sec:symmetry_suN} for the color structure in the density matrix.
}, namely $\rho_s(t) \equiv \rho_{ij,kl}(t)P^{(s)}_{ij,kl}$ and $\rho_o(t) \equiv \rho_{ij,kl}(t)P^{(o)}_{ij,kl}$, one can derive a closed equation for density matrix of the color-diagonal form
\begin{align}
\rho=\begin{pmatrix}
\rho_s & 0 \\ 0 & \rho_o
\end{pmatrix}.
\end{align}
The Lindblad operators $V_{\bm k}^a$ are now split into the four types
\begin{subequations}
\begin{align}
V^+_{\bm k} &= \left[e^{i\bm k\cdot\bm r/2} - e^{-i\bm k\cdot\bm r/2}\right]\sqrt{C_F}\begin{pmatrix}0 & 0 \\ 1 & 0\end{pmatrix},\\
V^-_{\bm k} &= \left[e^{i\bm k\cdot\bm r/2} - e^{-i\bm k\cdot\bm r/2}\right]\sqrt{\frac{1}{2N_c}}\begin{pmatrix}0 & 1 \\ 0 & 0\end{pmatrix},\\
V^d_{\bm k} &= \left[e^{i\bm k\cdot\bm r/2} - e^{-i\bm k\cdot\bm r/2}\right]\sqrt{\frac{N_c^2-4}{4N_c}}\begin{pmatrix}0 & 0 \\ 0 & 1\end{pmatrix},\\
V^f_{\bm k} &= \left[e^{i\bm k\cdot\bm r/2} + e^{-i\bm k\cdot\bm r/2}\right]\sqrt{\frac{N_c}{4}}\begin{pmatrix}0 & 0 \\ 0 & 1\end{pmatrix}.
\end{align}
\end{subequations}
The first two operators can be combined into one Lindblad operator
\begin{align}
V^{s\leftrightarrow o}_{\bm k}
= \left[e^{i\bm k\cdot\bm r/2} - e^{-i\bm k\cdot\bm r/2}\right]
\begin{pmatrix}0 & \sqrt{\frac{1}{2N_c}} \\ \sqrt{C_F} & 0\end{pmatrix}.
\end{align}
Dissipator expressed in the configuration space is obtained similarly
\begin{subequations}
\begin{align}
V^+_{\bm x} &= \left[\delta\left(\bm x- \frac{\bm r}{2}\right) - \delta\left(\bm x+\frac{\bm r}{2}\right)\right]\sqrt{C_F}\begin{pmatrix}0 & 0 \\ 1 & 0\end{pmatrix},\\
V^-_{\bm x} &= \left[\delta\left(\bm x- \frac{\bm r}{2}\right) - \delta\left(\bm x+\frac{\bm r}{2}\right)\right]\sqrt{\frac{1}{2N_c}}\begin{pmatrix}0 & 1 \\ 0 & 0\end{pmatrix},\\
V^d_{\bm x} &= \left[\delta\left(\bm x- \frac{\bm r}{2}\right) - \delta\left(\bm x+\frac{\bm r}{2}\right)\right]\sqrt{\frac{N_c^2-4}{4N_c}}\begin{pmatrix}0 & 0 \\ 0 & 1\end{pmatrix},\\
V^f_{\bm x} &= \left[\delta\left(\bm x- \frac{\bm r}{2}\right) + \delta\left(\bm x+\frac{\bm r}{2}\right)\right]\sqrt{\frac{N_c}{4}}\begin{pmatrix}0 & 0 \\ 0 & 1\end{pmatrix},
\end{align}
\end{subequations}
or the first two combined into one
\begin{align}
V^{s\leftrightarrow o}_{\bm x} &= \left[\delta\left(\bm x- \frac{\bm r}{2}\right) - \delta\left(\bm x+\frac{\bm r}{2}\right)\right]
\begin{pmatrix}0 & \sqrt{\frac{1}{2N_c}} \\ \sqrt{C_F} & 0\end{pmatrix}.
\end{align}
A distinct feature here is that the $V^+_{\bm x}$ and $V^-_{\bm x}$ (or $V^{s\leftrightarrow o}_{\bm x}$) are not any more hermitian.
Therefore, one cannot derive an equivalent stochastic potential description for the master equation in the singlet-octet basis.
Physically, the transition rates from the singlet to an octet state and those from an octet to the singlet are all the same (in the recoilless limit as adopted here).
Thus the rate from the singlet to octet in total is ($N_c^2-1$)-times larger than the opposite reaction and as a result $V^{s\leftrightarrow o}_{\bm x}$ is not hermitian.

\subsection{In the dipole limit}\label{app:lindblad_dipole}
Here we derive the stochastic potential in the dipole limit \eqref{eq:hamiltonian_dipole}.
First, expand the noise field $\Theta(\bm r,t)$ up to the second order in $r$:
\begin{align}
\label{eq:theta_dipole_naive}
&\Theta(\bm{r},t) =\theta^a(\bm{R}+\frac{\bm{r}}{2},t)(t^a\otimes 1)
-\theta^a(\bm{R}-\frac{\bm{r}}{2},t)(1\otimes t^{a*}) \nonumber \\
&\quad\quad\quad \simeq \left[\theta^a(\bm R, t) + \frac{r_ir_j}{8}\partial_i\partial_j\theta^a(\bm R, t)\right]\left(t^a\otimes 1 - 1\otimes t^{a*}\right) \nonumber\\
&\quad\quad\quad\quad + \frac{r_i}{2}\partial_i\theta^a(\bm R, t) \left(t^a\otimes 1 + 1\otimes t^{a*}\right) + \cdots.
\end{align}
The reason why we expand up to the second order is as follows:
The zero-th order noise $\Theta(\bm{r},t) \simeq\theta^a(\bm R, t)\left(t^a\otimes 1 - 1\otimes t^{a*}\right)$ is just a global color transformation and can be neglected if we are interested only in the singlet or total octet quantities.
The first order term contributes to the master equation only through the averages of their product $\propto r^2$, so that, for consistency, we need to expand $\Theta(\bm r,t)$ up to the second order.
The only contribution of the second order term is through its product with the zero-th order term.
However, such contribution turns out to cancel each other after projecting the density matrix onto the singlet and octet spaces.
So, as long as we measure singlet or total octet quantities, we can simply approximate
\begin{subequations}
\begin{align}
\label{eq:theta_dipole}
&\Theta(\bm{r},t) \simeq\frac{r_i}{2}\partial_i\theta^a(\bm R, t) \left(t^a\otimes 1 + 1\otimes t^{a*}\right),\\
&\langle \partial_i\theta^a(\bm R, t)\partial_j\theta^b(\bm R, t')\rangle = -\frac{\bm\nabla^2 D(0)}{3}\delta(t-t')\delta^{ab}\delta_{ij}.
\end{align}
\end{subequations}
Identifying $f^a_i(t)\equiv \partial_i\theta^a(\bm R, t)$ derives Eq.~\eqref{eq:hamiltonian_dipole}.
Note that with $D(\bm x)$ in Eq.~\eqref{eq:stoch_pot_pert}, its curvature at the origin $\bm\nabla^2 D(0)<0$ diverges.
This is because Eq.~\eqref{eq:stoch_pot_pert} is obtained in the soft approximation, which is applicable for $x\sim 1/gT$.
To get a finite value, one needs to improve the calculation for $x\sim 1/T$.
Physically, this quantity is related to the heavy quark momentum diffusion constant $\kappa = C_F\bm\nabla^2 D(0)/3$ and is obtained in the weak coupling expansion \cite{Moore:2004tg, CaronHuot:2007gq, CaronHuot:2008uh} as
\begin{align}
\kappa_{\rm NLO} &= \frac{C_F g^4 T^3}{18\pi}\left[\begin{aligned}
&N_c\left(\ln\frac{2T}{m_D} + \xi \right) \\
&+ \frac{N_f}{2}\left(\ln\frac{4T}{m_D} + \xi \right)
+\frac{N_cm_D}{T}C
\end{aligned}
\right], \\
\xi &= \frac{1}{2} - \gamma_E + \frac{\zeta'(2)}{\zeta(2)}\simeq -0.64718, \quad
C %= 1.4946 + \frac{21}{8\pi} 
\simeq 2.3302. \nonumber
\end{align}
By lattice QCD simulations, it is evaluated as $\kappa/T^3 \sim \mathcal O(1)$ for $T\sim (1\text{-}2) T_c$ \cite{Banerjee:2011ra, Francis:2015daa, Brambilla:2020siz}.
If we model $D(\bm x)$ by Eq.~\eqref{eq:stoch_pot_model}, we can get the small-$r$ limit without any difficulty.

Next, expand the Debye screened potential $V(\bm r)$ up to the second order in $r$, which is however not trivial.
The small-$r$ expansion of the Debye screened potential $V(\bm r)$ starts from the singular Coulomb part followed by a linear term $\propto r$ (and unphysical constant shift), but the correction to the Coulomb potential must start from $\propto r^2$ in the dipole limit, as is the case for potential NRQCD (pNRQCD) analysis.
This difference is due to the soft approximation employed to obtain the Debye screened potential.
Indeed, in the framework of NRQCD, we can trace back to the definition of $V(\bm r)$ in terms of a gluonic correlator, isolate the Coulomb singularity arising from the vacuum fluctuations, and then take the small-$r$ limit for the finite temperature contributions.
In this way, we obtain for small $r$\footnote{
There is a minor difference in the color structure between the potential obtained from NRQCD in the small-$r$ limit and the potential from pNRQCD with the dipole interaction \cite{Akamatsu:2020ypb}.
The octet potential from thermal fluctuation is $(-1/2N_c)\cdot(\lambda r^2 /2C_F)$ for the former and is $[(N_c^2-2)/4N_c]\cdot(\lambda r^2 /2C_F)$ for the latter with an opposite sign.
}
\begin{align}
\label{eq:potential_dipole}
V(\bm r) \simeq -\frac{\alpha_s}{r} + \frac{\lambda}{2C_F}r^2.
\end{align}
In the weak coupling expansion, thermal dipole self-energy coefficient $\lambda$ is obtained \cite{Eller:2019spw} as
\begin{align}
\lambda_{\rm NLO} = -2\zeta(3)C_F\left(\frac{4}{3}N_c + N_f\right)\alpha_s^2 T^3 + \frac{\alpha_sC_Fm_D^3}{3},
\end{align}
which can be extracted from \cite{Brambilla:2008cx}.
Since the Debye screened potential \eqref{eq:stoch_pot_pert} or \eqref{eq:stoch_pot_model} is not smoothly extrapolated to \eqref{eq:potential_dipole}, for the purpose of investigating the validity of small-$r$ limit, it would enough to take this limit for the noise term only.

The stochastic potential in the small-$r$ limit is equivalent to the following Lindblad equation
\begin{subequations}
\begin{align}
\label{eq:Lindblad_dipole}
&\frac{\partial}{\partial t}\rho(t)
= -i\left[\frac{\bm p^2}{M} + V(\bm r)(t^a\otimes t^{a*}), \rho\right] \\
&\quad +\frac{\kappa}{C_F}\left(C_i^a\rho C_i^{a\dagger}
- \frac{1}{2}C_i^{a\dagger}C_i^a\rho - \frac{1}{2}\rho C_i^{a\dagger}C_i^a\right), \nonumber\\
&C^a_i = \frac{r_i}{2} (t^a\otimes 1 + 1\otimes t^{a*}).
\end{align}
\end{subequations}
Similarly to the previous section, the Lindblad operators in the singlet-octet basis are
\begin{align}
C^{s\leftrightarrow o}_i &= r_i\begin{pmatrix}0 & \sqrt{\frac{1}{2N_c}} \\ \sqrt{C_F} & 0\end{pmatrix},\quad
C^d_i = r_i\sqrt{\frac{N_c^2-4}{4N_c}}\begin{pmatrix}0 & 0 \\ 0 & 1\end{pmatrix}.
\end{align}
Furthermore, we can derive a closed equation for the density matrices with fixed angular momenta
\begin{subequations}
\begin{align}
\rho_{s/o} &= \begin{pmatrix}
\rho_{s/o}^{(0)} & 0 & 0 & \cdots \\
0 & \rho_{s/o}^{(1)} & 0 & \cdots \\
0 & 0 & \rho_{s/o}^{(2)} & \cdots \\
\vdots & \vdots & \vdots & \ddots
\end{pmatrix}, \\
\rho_{s/o}^{(\ell)} &\equiv \sum_{m=-\ell}^{\ell} \langle \ell, m|\rho_{s/o}|\ell, m\rangle
\end{align}
\end{subequations}
in the Lindblad form.
Each Lindblad operator is split into raising and lowering operators on the angular momentum ladder
\begin{subequations}
\begin{align}
C^{s\leftrightarrow o}_+ &= r\begin{pmatrix}0 & \sqrt{\frac{1}{2N_c}} \\ \sqrt{C_F} & 0\end{pmatrix}\otimes\sum_{\ell}\sqrt{\frac{\ell+1}{2\ell+1}}
|\ell+1\rangle\langle\ell|,\\
C^{s\leftrightarrow o}_- &= r\begin{pmatrix}0 & \sqrt{\frac{1}{2N_c}} \\ \sqrt{C_F} & 0\end{pmatrix}\otimes\sum_{\ell}\sqrt{\frac{\ell}{2\ell+1}}
|\ell-1\rangle\langle\ell|,\\
C^d_+ &= r\sqrt{\frac{N_c^2-4}{4N_c}}\begin{pmatrix}0 & 0 \\ 0 & 1\end{pmatrix}\otimes\sum_{\ell}\sqrt{\frac{\ell+1}{2\ell+1}}
|\ell+1\rangle\langle\ell|,\\ 
C^d_- &= r\sqrt{\frac{N_c^2-4}{4N_c}}\begin{pmatrix}0 & 0 \\ 0 & 1\end{pmatrix}\otimes\sum_{\ell}\sqrt{\frac{\ell}{2\ell+1}}
|\ell-1\rangle\langle\ell|.
\end{align}
\end{subequations}
Once again we stress that the Lindblad operators $C_i^{s\leftrightarrow o}$ and $C_{\pm}^{s\leftrightarrow o, d}$ are not hermitian.
Therefore, it is not possible to obtain the stochastic potential corresponding to the Lindblad equations for the density matrices whose color and/or angular momentum subspaces are projected.

\section{Charge and spinor conjugations in the gauge theories}\label{app:symmetries}
The charge conjugation in the SU($N_c$) gauge theory with fundamental fermions is defined for Dirac field $\psi$ and for gauge field $A_{\mu}\equiv A_{\mu}^a t^a$ as
\begin{align}
\psi^C \equiv C\bar \psi^T, \quad
A_{\mu}^C \equiv - A_{\mu}^T,
\end{align}
where $C=i\gamma^2\gamma^0$ in the Dirac representation adopted hereafter.
The two component Pauli spinors transform as
\begin{align}
\psi = \left(\begin{aligned}\varphi \\ \chi\end{aligned}\right), \quad
\varphi^C = i\sigma^2\chi^*, \quad \chi^C = -i\sigma^2\varphi^*.
\end{align}
Note that in the charge conjugation, the color components of the fermions are unchanged.
To be explicit, we can write the transformation as tensor product of spinor/spin and color spaces 
\begin{subequations}
\begin{align}
\psi^C &= \left[C\gamma^0\otimes I\right]\psi^*, \\
\varphi^C &= \left[i\sigma^2\otimes I\right]\chi^*, \quad \chi^C = \left[-i\sigma^2\otimes I\right]\varphi^*.
\end{align}
\end{subequations}

For $N_c=2$, one can utilize a property of Pauli matrices
\begin{align}
\sigma^2 t^a\sigma^2 = -(t^a)^T
\end{align}
to absorb the charge conjugation transformation $A^C_{\mu} = - A_{\mu}^T$ into an additional transformation of the Dirac field:
\begin{subequations}
\begin{align}
\bar\psi \gamma^{\mu}A_{\mu}\psi &= -\bar\psi^C \gamma^{\mu}A^T_{\mu}\psi^C = \bar\psi^S \gamma^{\mu}A_{\mu}\psi^S, \\ 
\psi^S &= \left[C\gamma^0\otimes e^{i\eta} i\sigma^2\right]\psi^*, \quad
A_{\mu}^S = A_{\mu}.
\label{eq:spinorconj_dirac}
\end{align}
\end{subequations}
Here we include an arbitrary phase factor $e^{i\eta}$.
This is the spinor conjugation in the gauge theory.
The Pauli spinors transform as
\begin{subequations}
\begin{align}
\varphi^S &= \left[i\sigma^2 \otimes e^{i\eta} i\sigma^2\right]\chi^*, \\
\chi^S &= \left[-i\sigma^2 \otimes e^{i\eta} i\sigma^2\right]\varphi^*.
\end{align}
\end{subequations}
Since the gauge field is invariant under the spinor conjugation, $S$ restricted to each fermion species is still a symmetry; $S$ acting only for heavy quark is the symmetry we discussed in the main text.

Interestingly, $S^2 = -1$ for each fermion field.
Since $S$ is a multiplicative quantum number, for a system with heavy quark-antiquark pair $S^2 = (-1)^2 = 1$ so that $S=\pm 1$.
For such system, $S$ is determined by
\begin{align}
\label{eq:S_QQbar}
S = (-1)^{\ell + s + c},
\end{align}
where $\ell=0,1,2,\cdots$ is orbital angular momentum, $s=0,1$ for spin singlet and triplet, and $c=0,1$ for color singlet and triplet.

It is worthwhile to point out that the spinor conjugation shares its algebraic properties with $G$-parity $G = C e^{i\pi I_2}$, where $I_2$ is a generator of isospin SU(2) group \cite{Lee:1956sw, goebel1956selection}.
Indeed, we can also write $S = C e^{i\pi T_2}$, where $T_2$ is a generator of color SU(2) group and $\eta=0$ is assumed.
For charge neutral systems, $G$ is determined similarly to \eqref{eq:S_QQbar} by $G=(-1)^{\ell + s + I}$, where $I$ is the total isospin number.
The difference is the transformation of the force carriers, i.e. gauge field $A_{\mu}$ and pion field $\Pi \equiv \pi^a t^a$:
\begin{subequations}
\begin{align}
A^C_{\mu} &= -A_{\mu}^T, \quad A_{\mu}^S = e^{i\pi \sigma^2/2}A^C_{\mu}e^{-i\pi \sigma^2/2} = A_{\mu},\\
\Pi^C &= \Pi^T, \quad \Pi^G = e^{i\pi \sigma^2/2}\Pi^C e^{-i\pi \sigma^2/2} = -\Pi.
\end{align}
\end{subequations}
Contrary to the gluons, the pions are $G$-odd leading to selection rules for annihilations of nucleon-antinucleon systems into pions \cite{Lee:1956sw, goebel1956selection}.
Also, if we derive open system description for a nucleon-antinucleon system in a pion gas environment, the $G$-parity is not an event-by-event symmetry because pions are $G$-odd.

In the main text, we ignored the spin degrees of freedom.
It corresponds to assigning the following transformations for charge and spinor conjugations
\begin{subequations}
\begin{align}
\varphi^C &= \chi^*, \quad \chi^C = \varphi^*, \\
\varphi^S &= e^{i\eta} i\sigma^2\chi^*, \quad \chi^S = e^{i\eta}i\sigma^2\varphi^*.
\end{align}
\end{subequations}
The wave function of the heavy quark-antiquark pair with color $i$ and anti-color $j$ is defined by
\begin{align}
\Psi_{ij}(\bm r,t) \equiv \langle \varphi_i(\bm R+\frac{\bm r}{2})\chi_j^{\dagger}(\bm R-\frac{\bm r}{2})|\Psi(t)\rangle.
\end{align}
It transforms under charge and spinor conjugations as
\begin{align}
\label{eq:Psi_C}
\Psi_{ij}^C(\bm r,t) &= \langle \varphi_i^C(\bm R+\frac{\bm r}{2})\chi_j^{C\dagger}(\bm R-\frac{\bm r}{2})|\Psi(t)\rangle \nonumber \\
&= \langle \chi_i^{\dagger}(\bm R+\frac{\bm r}{2})\varphi_j(\bm R-\frac{\bm r}{2})|\Psi(t)\rangle \nonumber \\
&= -\Psi_{ji}(-\bm r, t)
\end{align}
and
\begin{align}
\label{eq:Psi_S}
\Psi_{ij}^S(\bm r,t) &= \langle \varphi_i^S(\bm R+\frac{\bm r}{2})\chi_j^{S\dagger}(\bm R-\frac{\bm r}{2})|\Psi(t)\rangle \nonumber \\
&= \epsilon_{il}\epsilon_{jk}\langle \chi_l^{\dagger}(\bm R+\frac{\bm r}{2})\varphi_k(\bm R-\frac{\bm r}{2})|\Psi(t)\rangle \nonumber \\
&= -\epsilon_{il}\epsilon_{jk}\Psi_{kl}(-\bm r, t) .
\end{align}

\bibliography{SU3stochastic_v8}

\end{document}